\documentclass[fleqn,usenatbib]{mnras}

\usepackage{newtxtext,newtxmath}
\usepackage{comment, cleveref}

\usepackage[T1]{fontenc}
\usepackage{array}

\usepackage{booktabs} 
\usepackage{longtable}
\usepackage{tabularx}
\usepackage{adjustbox}  
\usepackage{orcidlink}
\usepackage{hyperref}

\DeclareRobustCommand{\VAN}[3]{#2}
\let\VANthebibliography\thebibliography
\def\thebibliography{\DeclareRobustCommand{\VAN}[3]{##3}\VANthebibliography}

\usepackage{graphicx}	
\usepackage{amsmath}	
\newcommand{\code}[1]{\texttt{#1}}

\title{The promise of self-supervised and active learning for Strong Lens discovery: \texttt{Astronomaly} applied to KiDS}

\author[M. Grespan et al.]{
Margherita Grespan\orcidlink{0000-0002-8213-0217},$^{1,2}$\thanks{E-mail: margherita.grespan@physics.ox.ac.uk}
Aprajita Verma\orcidlink{0000-0002-0730-0781},$^{1}$
Michelle Lochner \orcidlink{0000-0003-2221-8281},$^{3}$
Koketso Mohale\orcidlink{0000-0001-6551-4107},$^{3}$
Verlon Etsebeth\orcidlink{0000-0001-9513-6442},$^{3}$
\newauthor
Duncan Bowden\orcidlink{0009-0008-6114-140}$^{1}$\\
$^{1}$ Sub-department of Astrophysics, University of Oxford, Denys Wilkinson Building, Keble Road, Oxford OX1 3RH, UK\\
$^{2}$ National Centre for Nuclear Research, ul. Pasteura 7, 02-093 Warsaw, Poland\\
$^{3}$ Department of Physics and Astronomy, University of the Western Cape, Bellville, Cape Town 7535, South Africa
}

\date{Accepted XXX. Received YYY; in original form ZZZ}

\pubyear{\the\year{}}

\begin{document}
\label{firstpage}
\pagerange{\pageref{firstpage}--\pageref{lastpage}}
\maketitle

\begin{abstract}
Strong gravitational lenses (SGLs) are rare systems whose discovery currently relies primarily on supervised machine learning methods trained on large simulated datasets. We present the first application of Astronomaly:PROTEGE to SGL discovery, demonstrating that a human-in-the-loop active learning framework can efficiently identify lenses in large imaging surveys without the need for simulated training data. We consider a sample of 3.7 million bright galaxies from the Kilo-Degree Survey (KiDS) DR4. Feature representations are extracted using a convolutional neural network pre-trained on the ImageNet dataset and subsequently fine-tuned on KiDS data using the self-supervised Bootstrap Your Own Latent (BYOL) framework.  Within the embedding of these representations, the active learning loop of \texttt{Astronomaly} iteratively selects the most informative systems for expert inspection. A total of 3,000 objects are inspected across multiple rounds, yielding 34 high-quality (grade A/B) SGL candidates. On the basis that these systems occupy similar regions in the learned feature space, we expand this sample through nearest-neighbour similarity analysis. Including the active learning discoveries, we identify a total of 140 grade A/B candidates and more than 1,000 additional lower-confidence systems (grade C). Among the A/B candidates, 81 are newly identified, while $\sim$22\% of previously known grade A/B KiDS lenses are recovered. These results demonstrate strong potential for next-generation surveys such as \textit{Euclid}, \textit{Roman}, and Rubin’s Legacy Survey of Space and Time. With approximately 60\% of the high-quality candidates newly reported, this approach complements supervised methods by reducing reliance on simulations and enabling the discovery of a diverse population of SGLs.
\end{abstract}


\section{Introduction} 
\label{sec:introduction_anomaly}
Strong gravitational lensing occurs when a massive foreground object foreground object, typically an early-type galaxy or galaxy cluster, lies close to the line of sight to a distant background source, such as a galaxy or quasar. Under these conditions of near-perfect alignment, the gravitational potential of the lens produces multiple images, arcs, or Einstein rings \citep[see][for a review]{Treu_2010}. Because this configuration is highly specific, strong gravitational lenses (SGLs) are intrinsically rare, with probabilities on the order of one per $\sim10^3$ massive galaxies \citep{Collett_2015}. 
SGLs are powerful astrophysical and cosmological probes. They enable direct, model-independent measurements of the total mass distribution in galaxies - probing both baryonic and dark matter, as well as their evolution \citep[e.g.][]{Auger_2009,Barnabe_2012}. They further constrain the stellar initial mass function and inner density profiles of massive early-type galaxies \citep[e.g.][]{Koopmans_2006, Sonnenfeld_2019_sugohi, Geng2024}, and provide powerful tests of dark matter physics \citep{Vegetti_2010, vegetti_2023_review, Gilman_2024}. In time-variable systems, SGLs offer precise cosmographic measurements, including constraints on $H_0$ and dark energy models \citep[e.g.][]{Suyu_2013, sharma_2023, Li_2023, Treu_2023, birrer_2024_review}. For a comprehensive review of galaxy–galaxy lensing applications, see \citet{Shajib_2024}.

Despite more than two decades of systematic searches, only a few thousand candidates have been identified, and an even smaller subset has been spectroscopically confirmed \citep[e.g.][]{Bolton_2008, More_2016, Tran_2022, barone2026agel}. However, this landscape is poised to change. Forthcoming wide-field surveys such as those from the Vera C. Rubin Observatory's Legacy Survey of Space and Time (LSST) \citep{Ivezic_2019}, \textit{Euclid}  \citep{laureijs2011euclid}, and the Nancy Grace Roman Space Telescope \citep{Mosby_2020} are expected to observe millions of massive galaxies, raising the prospect of discovering over $10^5$ SGLs \citep{Collett_2015}. Early \textit{Euclid} results already report surface densities of up to $\sim10$ lenses per square degree \citep{nagam_2025euclidfindingstronggravitational, euclidcollaboration_2025_Walmsley, Acevedo_Barroso_2025}. In comparison, searches in Hyper Suprime Cam \citep[HSC,][]{Hiroaki_2018}, which reaches LSST-like depth \citep{Ivezic_2019}, have identified approximately 1 SGL per square degree. This difference is primarily driven by \textit{Euclid}'s higher spatial resolution and near-infrared sensitivity, which enhance the detectability of small-separation and high-redshift lens systems, highlighting the significant increase in discovery potential offered by next-generation surveys.

Identifying SGLs in wide-area survey scale remains a major challenge, with supervised machine learning methods now serving as the primary first-stage filter for candidate discovery and contamination suppression. While these methods have achieved notable success, leading to comparative studies \citep[e.g.][]{gonzalez2025doesmachinelearningwork,canameras2023holismokes, More_2024, pearcecasey_2024} and dedicated open data challenges \citep[e.g.][]{Metcalf_2019, Bom_2022}, their performance typically deteriorates in two well-known scenarios: (i) when detecting previously unseen or intrinsically rare phenomena, and (ii) when the training data differ from the target domain, a mismatch known as domain shift or the generalisation gap \citep{quinonero_2022_datashift, Zhang_domainshift_2021}.

These limitations are particularly severe for SGL searches. Because SGLs are extremely rare, even high validation performance often fails to translate to survey conditions, where the vast majority of objects are non-lenses and false positives dominate—a manifestation of the base-rate fallacy. The intrinsic rarity of SGLs also makes it difficult to assemble large, diverse, and representative labelled datasets from real observations alone. As a result, simulations are widely used \citep[e.g.][]{Metcalf_2019}, but they inevitably differ from real survey data, reducing model accuracy and reliability \citep{CiprijanoviC_2021, grespan2024teglie, More_2024}. 
While surveys such as \textit{Euclid} are expected to significantly increase the number of known lenses, these samples are survey-specific, which can limit the direct transferability of models trained on one dataset to another. In practice, cross-survey applications require explicit adaptation, as demonstrated by \citet{Thuruthipilly_2025}, and can lead to degraded performance if not properly fine-tuned.
Moreover, real SGLs can exhibit morphologies that are not represented in the training data, making them effectively “unseen” to the model. As a result, domain gaps between simulated examples and real survey images are difficult to eliminate.

Hybrid strategies, such as `painting' simulated arcs onto real galaxies \citep{More_2015, Petrillo_2017, Jacobs_2017, rojas2021strong, Canameras_2020, Canameras_2021} or using Active machine learning to build the training set using real examples \citep{gonzalez_2025,saamie_2026}, can partially mitigate this issue.

One promising avenue for reducing the FP is the adoption of foundation models \citep{bommasan_2021}, large neural networks pre-trained on extensive and heterogeneous astronomical datasets and subsequently fine-tuned for specific downstream tasks \citep{pearcecasey_2024}. For example, \texttt{Zoobot} \citep{walmsley_2023zoobot, walmsley2024scaling} has been integrated into the \textit{Euclid} Strong Lensing Discovery Engine as one of the detection models, where it operates within a multi-stage detection pipeline. In this framework, initial candidate samples are first filtered through citizen science classifications, followed by expert inspection, and subsequently refined using lens modelling \citep{euclidcollaboration_2025_Walmsley}. 
However, as a supervised approach trained for specific datasets, its performance may be limited when previously unseen lens configurations, and can be affected by domain shift when applied to new surveys.

Another option is to work directly with survey data rather than relying on synthetic training sets. A practical way to achieve this is through active learning \citep[AL, see][for in-depth reviews]{settles_2009_active, settles_2011}.
In AL, the model identifies informative unlabeled images for review, a human expert assigns labels, and the model is retrained after each labeled batch before choosing the next set of candidates.
Research on AL in astronomy is rapidly expanding, with applications ranging from transient to quasar identification \citep{Ishida_2021, green_2025}, and a growing emphasis on active anomaly detection \citep{gomez_2025_anomalymatch, oryan_2025_HST, Anderson_2002, kornilov2025coniferest}. Recent work has demonstrated the power of these approaches for large-scale searches, for example, identifying anomalies - including SGLs - within nearly $10^8$ Hubble Space Telescope images using the \texttt{AnomalyMatch} framework \citep{gomez_2025_anomalymatch}. Here, the term “anomaly’’ refers to rare or atypical astrophysical objects, often providing informative constraints on galaxy evolution and cosmology. 

To implement the AL loop, in this work we adopt \texttt{Astronomaly: PROTEGE} \citep{lochner_2024, Lochner_2021}. Although originally developed for general active anomaly detection, the framework can be used without any anomaly detection algorithm (e.g., Isolation Forest, \citealt{Liu_2008}), enabling a purely AL-driven selection, an approach well suited to SGL searches, where true SGL tend not to appear as statistical outliers. 
The \texttt{PROTEGE} update \citep{lochner_2024} adopts a Gaussian Process \citep[GP;][]{Rasmussen2004} regression–based prioritisation scheme, following the approach of \citet{Walmsley_2022_GP}, to predict human scores and rank candidates. In this context, the GP learns a mapping between the feature representation of each object and the scores assigned by the user during the active learning loop, effectively modelling the user’s notion of “interestingness.” This prediction is then used to guide exploration within the embedding space, selecting new objects based on GP-driven uncertainty estimates.

A key requirement for AL is a compact and informative representation of the data. Self-supervised learning (SSL) provides a framework for learning such representations directly from the data itself, without the need for large sets of manual labels, by defining surrogate tasks that encourage the model to capture meaningful structure in the inputs; refer to \citet{gui_2024_SSL, SSL_liu_2023} for recent reviews on self-supervised methods. SSL has recently gained traction in astronomy, demonstrating strong performance in learning robust, survey-specific representations from imaging data \citep{walmsley_2022_SSL, Slijepcevic_2024, Mohale_2024, riggi_2024}.
Unsupervised approaches have also been explored in the context of SGL searches, either by augmenting training sets with GAN-generated images \citep{Sheng_2022_hunt, Keerthi_2023} or by refining models through self-supervised pretraining and similarity-based retrieval methods \citep{Stein_2022}. Among SSL methods, the Bootstrap Your Own Latent (\texttt{BYOL}) framework \citep{Grill_2020} has proven particularly effective for astronomical imaging, producing high-quality feature representations suitable for downstream tasks \citep[e.g.][]{walmsley_2022_SSL, Mohale_2024}.

Recent work has demonstrated that \texttt{Astronomaly} can be applied at survey scale \citep{Verlon_2024}. In that study, \texttt{Astronomaly} \citep{Lochner_2021} - combined with representations extracted from the data with \texttt{Zoobot} - was used on the Dark Energy Camera Legacy Survey \citep[DECaLS,][]{Decals_2019} imaging to identify unusual astrophysical systems. The approach successfully recovered a variety of rare objects, including eight SGL candidates and numerous galaxy mergers, but with a clear bias toward mergers. Given that mergers are far more common than SGLs, they tend to dominate anomaly-search rankings. 

\citet{Mohale_2024} demonstrated the use of \code{BYOL}-based representations within \texttt{Astronomaly:PROTEGE} in optical galaxy images, highlighting the benefits of deep self-supervised feature learning for large-scale searches. More recently, \cite{etsebeth2026targetedmachinelearningapproach} combined \code{BYOL} features with \texttt{Astronomaly:PROTEGE} to identify diffuse radio emission with minimal human labelling effort. 
However, prior to this work, the performance of this framework when targeting a single, extremely rare class of object has not yet been systematically explored. Building on these developments, this work investigates the suitability of \texttt{Astronomaly} for survey-level identification of a single, specific class of anomaly, with a particular focus on SGLs.

The paper is organized as follows. Section~\ref{sec:data_anomaly} describes the dataset and pre-processing, Section~\ref{sec:methodology} outlines the methodology, Section~\ref{sec:results} presents the results, Section~\ref{sec:discussion} discusses their implications, and Section~\ref{sec:conclusions} summarises our conclusions.

\section{data, selection and preprocessing}
\label{sec:data_anomaly}

This work uses the fourth data release (DR4) of the Kilo-Degree Survey\footnote{\url{https://kids.strw.leidenuniv.nl/DR4/index.php}} \citep[KiDS,][]{Kuijken_2019}, which covers approximately 1,000\,deg$^{2}$ of sky in the optical $ugri$ bands. The $5\sigma$ limiting AB magnitudes (measured in $2''$ apertures) are $24.23 \pm 0.12$, $25.12 \pm 0.14$, $25.02 \pm 0.13$, and $23.68 \pm 0.27$ in $u$, $g$, $r$, and $i$, respectively, where the uncertainties represent the RMS scatter from tile to tile. The mean seeing in the $r$ band is $0.70''$.

\subsection{Selection cuts}

\label{sec:sel_cuts}

Following earlier searches  \citep{Petrillo_2017, Petrillo_2018, Li_2020, Li_2021}, we focus on bright galaxies (BGs) - galaxies with the highest lensing cross-section \citep{Oguri_2010} - selected using the following criteria:

\begin{itemize}
    \item \code{SExtractor} \citep{Bertin_1996} $r$-band \code{FLAG} $< 4$, excluding objects affected by corrupted photometry, saturation, or blending.
    \item \code{IMA\_FLAGS} = 0, removing sources located in compromised image regions.
    \item Kron-like $r$-band magnitude \code{MAG\_AUTO} $\leq 21$\,mag, maximising the lensing cross-section \citep{Schneider_1992}.
    \item KiDS star–galaxy separation parameter \code{SG2DPHOT} = 0, selecting galaxy-like sources. This parameter takes the value 1 for stars, 2 for unreliable sources, 4 for stellar objects based on classification criteria, and 0 for non-stellar objects (see \citealt{La_Barbera_2008, deJong2015, Kuijken_2019} for details).
\end{itemize}
With these cuts, we obtain a sample of 3\,695\,703 galaxies. 

We generate $64 \times 64$ pixel cutouts, corresponding to approximately $13'' \times 13''$. RGB images are constructed using the $r$, $i$, and $g$ bands, combined following the colour scheme proposed in \citet{Lupton_2004}.

\subsection{Evaluation sets}

\subsubsection{Previously discovered lens candidates}
\label{sec:prev_disc}
KiDS provides an excellent testbed for machine learning techniques and has been among the early surveys in which such methods have been extensively applied to strong lens discovery. As a result, several catalogues of SGL candidates are available in the literature. 

To assess our results, we make use of previously identified SGL candidates, restricting the analysis to systems detected within KiDS to ensure a fair comparison with other lens-finding approaches. Systems discovered in other surveys that fall within the KiDS footprint are not necessarily detectable in KiDS data due to differences in depth, image quality, and point spread function.

We consider the 165 candidates in KiDS DR4 identified by previous KiDS collaboration searches \citep{Petrillo_2017, Petrillo_2018, Li_2020}, as reported on the KiDS DR4 website\footnote{\url{https://kids.strw.leidenuniv.nl/DR4/hqlenses.php}}, together with the 210 candidates presented in our previous work \citet[][hereafter G+24]{grespan2024teglie}. The G+24 sample is based on labels assigned through expert visual inspection of candidates identified by a machine learning model in a targeted SGL search within the 221 deg$^2$ KiDS-GAMA overlap region \citep{Driver_2009}. In addition to SGL candidates, the G+24 dataset includes a large population of galaxies labelled as non-lenses. Applying the same BG selection used in this work (Sec.~\ref{sec:sel_cuts}) to all the labelled objects in G+24 we obtain a subset of 40,738 galaxies (the G+24 sample).
We hereafter refer to the KiDS collaboration candidates as the ``KiDS'' sample, the G+24 candidates as ``TEGLIE''\footnote{Transformer Encoders as strong Gravitational Lens finders In the kilo-degreE survey}, and the combined set as the ``Known'' sample.

It is important to note that these candidates originate from a heterogeneous set of searches carried out at different times and with different machine learning models. Over this period, both machine learning models and human inspection criteria have evolved, with classifiers improving in performance and grading standards becoming increasingly stringent. Furthermore, SGL inspection and grading remain inherently subjective \citep{Rojas_2023}, as they depend on individual assessors and varying classification schemes. 

To ensure consistency throughout this work, all candidates are visually inspected by the same three authors (MG, AV, DB), and a uniform classification scheme is applied across the entire dataset. Each candidate is assigned one of four labels based on visual assessment:

\begin{itemize}
    \item \texttt{A}: secure SGL, exhibiting clear lensing features with no additional information required;
    \item \texttt{B}: probable lens, showing convincing lensing features but requiring additional information for confirmation;
    \item \texttt{C}: possible lens, displaying features that could be explained without invoking gravitational lensing;
    \item \texttt{X}: non-lens.
\end{itemize}

The discrete labels are mapped to numerical scores, with grade~A assigned a value of 3, grade~B a value of 2, grade~C a value of 1, and grade~X a value of 0. For each object, the final score is defined as the average of the three independent classifications.

Grade boundaries are defined based on the distribution of average scores for the known candidates, selecting thresholds that best separate the different quality classes. Objects with $\langle s \rangle \geq 2.0$ are classified as grade~A, those with $1.0 \leq \langle s \rangle < 2.0$ as grade~B, and those with $0 < \langle s \rangle < 1.0$ as grade~C.
Our purpose in regrading is to identify systems of high and low confidence. For this reason, the grade boundaries are determined in a second step: the candidates are sorted by average score and inspected as mosaics. The thresholds are placed where the visual quality of the systems changes, rather than by mapping average scores back onto the original grade definitions. The reclassified grades should therefore be read as relative confidence levels within this work.

Although the Known systems were previously classified as high-quality candidates, the updated visual assessment assigns a substantial fraction to lower grades, as summarised in Table~\ref{tab:origin_grades}, reflecting the more conservative criteria adopted in this work.

\begin{table}
\centering
\caption{Number SGL candidates per origin and final visual grade. Candidates reported by multiple searches contribute to each corresponding origin.}
\label{tab:origin_grades}
\begin{tabular}{lrrrrr}
\toprule
Origin  & A & B & C & X & Total \\
\midrule
TEGLIE  &  4 & 35 & 135 & 36 & 210 \\
KiDS   & 12 & 65 & 78  & 10 & 165 \\
\midrule
Total   & 15 & 92 & 206 & 45 & 358 \\
\bottomrule
\end{tabular}
\end{table}

\subsubsection{Construction of evaluation samples}
\label{sec:label_data}

\texttt{Astronomaly} relies on iterative user feedback to refine candidate ranking during the active learning process. While this feedback is typically provided interactively, the availability of a labelled dataset allows the process to be reproduced by automatically supplying labels in place of user input. This is particularly useful for optimising the active learning configuration and assessing performance in a controlled setting, typically on a subset of the full dataset.

The G+24 sample was originally validated using a discrete grading scheme, with grade~0 denoting non-lenses (X), grade~1 corresponding to A candidates, grade~2 to B candidates, and grade~3 to systems exhibiting lens-like features but not consistent with strong gravitational lensing (i.e. intermediate between C and X).

In contrast, the active learning framework in \texttt{Astronomaly} requires user-provided scores on a continuous scale from 0 to 5, where 0 indicates no interest and 5 corresponds to the target class of interest (see Section~\ref{sec:AL}). To adapt the G+24 labels to this scheme, we apply a mapping in which objects graded as 1 are assigned a score of 5, those graded as 2 a score of 4, while grades 0 and 3 are retained unchanged.

In addition to the original G+24 labels, we also consider the updated classifications derived from our visual re-inspection (Table~\ref{tab:origin_grades}). These revised grades, expressed in terms of the A/B/C/X scheme adopted in this work, are mapped onto the active learning score range by assigning scores of 5, 4, 3, and 0 to A, B, C, and X candidates, respectively. This allows us to construct a second labelled dataset consistent with the active learning framework.

In this way, we obtain two evaluation samples: one based on the original G+24 classifications, and a second based on the updated labels introduced in this work, with both samples consistently mapped to the active learning score range.

\subsection{Pre-processing}
\label{sec:preprocessing}

\subsubsection{Resnet adaptation}
  
To ensure compatibility with the pre-trained ResNet  \citep{he_2015_resnet} architecture, each input RGB image is resized to \(224 \times 224\) pixels using the default \texttt{torchvision} interpolation (bilinear), matching the expected input dimensions of the model. Pixel values are scaled to the range \([0,1]\) and subsequently normalised on a per-channel basis using the ImageNet \citep{ImageNet} mean values \((0.485, 0.456, 0.406)\) and standard deviations \((0.229, 0.224, 0.225)\) for the red, green, and blue channels, respectively. This preprocessing matches the input distribution of the ImageNet dataset on which the network was originally trained. All transformations are implemented using the \texttt{torchvision} library \citep{Torchivision}.

\subsubsection{Sigma clipping}

Background structures and noise can affect feature extraction and degrade subsequent analysis. To mitigate this, we sigma-clip the images to isolate the SGL and its associated arcs while suppressing background fluctuations. A $2.5\sigma$ threshold is adopted, based on visual inspection of the Known sample in the G+24 dataset, as it effectively preserves faint lensing features while limiting residual background structure.

For mask generation, RGB images are converted to grayscale, which, based on visual inspection of known candidates, provides more reliable contours than the $r$-band alone. Sigma-clipped statistics (mean, median, and standard deviation) are computed on the grayscale images using \code{AstroPy} \citep{Astropy_2013, astropy_2018, Astropy_2022} to define the clipping threshold. The resulting contours are identified and applied to the original RGB images using \code{OpenCV} \citep{opencv_library}.

\section{Methodology}
\label{sec:methodology}

We explore an approach for the discovery of SGLs that combines self-supervised feature learning, active learning, and similarity search. The aim is to efficiently identify SGL candidates in large imaging datasets while minimising the need for extensive labelled training data.

Starting from KiDS multi-band imaging, RGB cutouts are generated for each galaxy. A ResNet-18 model, pre-trained on the ImageNet dataset (Sec.~\ref{sec:CNN}), is fine-tuned using the self-supervised \code{BYOL} \citep{Grill_2020} framework to learn a compact representation of the data without requiring labels (Sec.~\ref{sec:finetuning}). The fine-tuned network is then used to extract feature embeddings for all $\sim$3.7 million BGs in KiDS DR4. These embeddings are then reduced in dimensionality using principal component analysis \citep[PCA;][]{Pearson_1901}, enabling efficient storage and downstream analysis (Sec.~\ref{sec:feat_dim_red}).

The resulting feature space is explored using the \texttt{Astronomaly:Protege}\footnote{\url{https://github.com/MichelleLochner/astronomaly}} active learning framework, which iteratively ranks objects based on user-provided feedback (Sec.~\ref{sec:SGLfinding_astro}). At each iteration, the user inspects a subset of high-priority candidates and assigns scores reflecting their likelihood of being SGLs. These scores are used to update the Gaussian Process, which refines its ranking by balancing exploration of uncertain regions and exploitation of high-scoring areas in feature space (Sec.~\ref{sec:GP}). This human-in-the-loop process progressively concentrates the search on regions enriched in lens-like systems.

Objects receiving high scores are then subject to further visual inspection and graded by the authors MG, AV and DB following the scheme described in Sec.~\ref{sec:prev_disc}. The neighbourhood of these promising candidates in feature space is explored through a similarity search (Sec.~\ref{sec:sim_search}), enabling the identification of additional objects with comparable morphology that may not have been directly selected by the active learning model.
These newly identified candidates are subsequently re-inspected and graded following the same procedure. This combination of global (active learning) and local (similarity search) exploration improves the completeness of the candidate sample.

\begin{figure*}
    \centering
    \includegraphics[width=\textwidth,keepaspectratio]{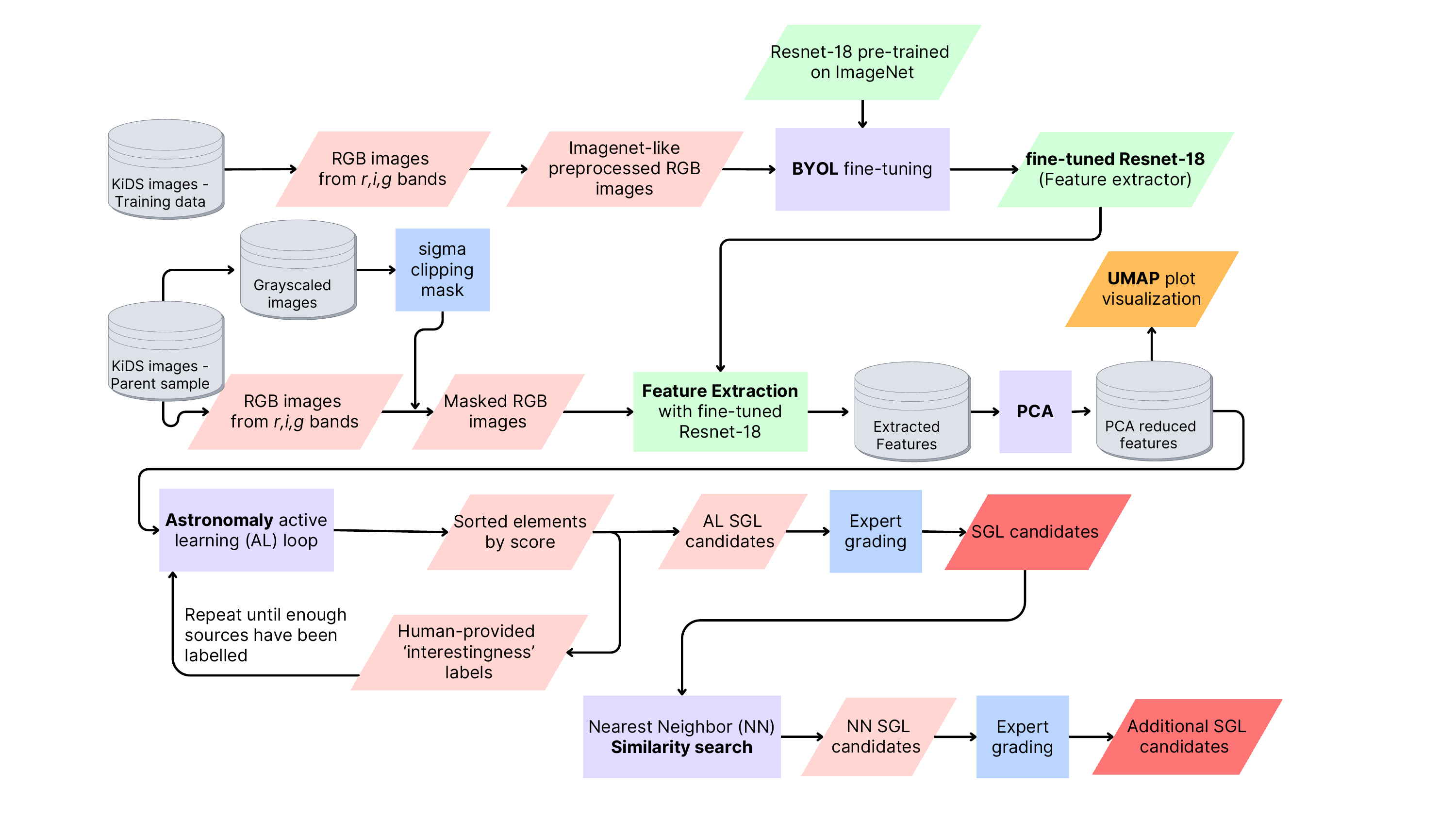}
    \caption{Flowchart illustrating the methodology: parallelograms represent inputs, rectangles actions, and stacked cylinders datasets.}
    \label{fig:flowchartbyol}
\end{figure*}

A schematic overview of the methodology is shown in Fig.~\ref{fig:flowchartbyol}.

We also make publicly available on GitHub \footnote{\url{https://github.com/margres/KiDS_lenses_astronomaly_protege}} the code developed for this work, including the adapted \texttt{Astronomaly:Protege} code, the BYOL fine-tuning and feature-extraction scripts, and the labelling interface used during nearest neighbour inspection.

\subsection{CNNs as Feature Extractors}
\label{sec:CNN}

RGB images are represented as three-dimensional arrays (\(n \times n \times c\)), where \(n\) is the pixel dimension of the image, and \(c\) refers to the number of colour channels.
Most unsupervised algorithms encounter difficulties in handling such high-dimensional data.  To overcome this, they rely on simplified, lower-dimensional representations known as features. Feature extraction is thus an essential step to reduce data complexity while preserving information. 

Pre-trained models, particularly convolutional neural networks \citep[CNN;][]{Lecun}, are well suited for feature extraction \citep{Feature_extraction}, offering substantial reductions in computational cost and training time owing to large-scale prior training. However, such networks remain sensitive to the domain on which they were originally trained. Self-supervised fine-tuning has therefore gained increasing attention as a means of adapting a pre-trained model to a target dataset, allowing the network to retain general features learned from large-scale training while refining them to better capture the specific structures and characteristics of the new domain.

Following \citet{Mohale_2024}, which shows that CNNs initialised with pre-trained weights outperform those trained from random initialisation, we adopt a ResNet18 architecture \citep{he_2015_resnet} implemented in \texttt{torchvision}\footnote{\url{https://pytorch.org/vision/main/models/generated/torchvision.models.resnet18.html}} and initialised with ImageNet weights \citep{ImageNet}.

ResNet \citep{he_2015_resnet} is a deep CNN architecture designed to mitigate the vanishing gradient problem \citep{Bengio_1994_gradient, Glorot_2010_gradient} through residual connections. ResNet18, a lightweight variant comprising 18 layers, provides an effective balance between computational efficiency and representational capacity. Compared to deeper models such as ResNet50, ResNet101, or ResNet152, it is less computationally demanding while maintaining strong performance, making it well suited to this application.

\subsection{Fine-tuning with Self-Supervised Learning}
\label{sec:finetuning}

SSL aims to learn informative representations directly from unlabelled data by optimising auxiliary tasks that do not require manual annotations. In computer vision, many modern SSL approaches enforce consistency between different augmented views of the same image, encouraging the model to capture invariant and semantically meaningful features.

Contrastive methods \citep[e.g.][]{chen2020simple} achieve this by training the model to bring representations of positive pairs (two augmented views of the same image) closer together while pushing apart those of negative pairs (views from different images). While effective, this requires large batches or memory banks to ensure sufficient exposure to negative examples, and performance is sensitive to their selection.

\citet{Grill_2020} introduced \texttt{BYOL}, a non-contrastive alternative that removes the need for negative pairs entirely. Without them, however, the model risks representation collapse, where all inputs are mapped to the same representation. \texttt{BYOL} avoids this through an asymmetric architecture consisting of an online network and a momentum-updated target network: the online network is trained to predict the target network's representation of a differently augmented view of the same image, while the target network is updated as an exponential moving average of the online network weights rather than by gradient descent \citep{tian_2021}. This asymmetry is sufficient to prevent collapse without requiring negative pairs.

\begin{figure}
    \centering
    \includegraphics[width=\linewidth]{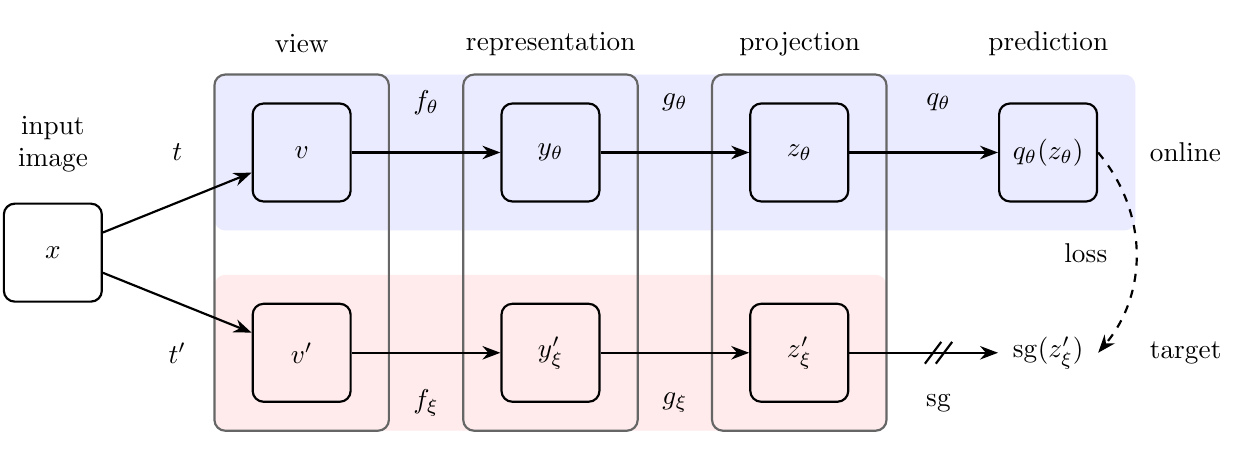}
    \caption{BYOL minimizes a similarity loss between $q_\theta(z_\theta)$ and $\text{sg}(z'_\xi)$, where $\theta$ are trainable weights, $\xi$ their exponential moving average, and $\text{sg}$ the stop-gradient. After training, only $f_\theta$ is retained, with $y_\theta$ used as the image representation. Image credit: \citet{Grill_2020}.  }
    \label{fig:byol}
\end{figure}

As illustrated in Fig.~\ref{fig:byol}, the framework employs two networks: an online network and a target network. Both receive different augmented views of the same input image $x$. The online network processes the first augmented view $t(x)$ through its encoder $f_\theta$, producing a representation $y_\theta$, while the target network processes a second augmented view $t'(x)$ through its encoder $f_\xi$, yielding the representation $y_\xi$. The online network representation $y_\theta$ is passed through a projection head $g_\theta$, yielding the projection $z_\theta$. This projection head, implemented as a fully connected neural network, maps the high-dimensional representation into a lower-dimensional space, simplifying the prediction task. Similarly, the target network applies its projection head $g_\xi$ to $y_\xi$, producing the projected representation $z'_\xi$. A predictor network $q_\theta$ is applied to $z_\theta$ to generate the prediction $q_\theta(z_\theta)$. The training objective minimises the mean squared error between $q_\theta(z_\theta)$ and the target projection $z'_\xi$. Gradients are not propagated through the target network, which is treated as a stop-gradient (\textit{sg}) operation. The target network parameters $\xi$ are updated as an exponential moving average of the online parameters $\theta$:
\[
\xi_i \leftarrow \tau \xi_{i-1} + (1 - \tau)\theta_i,
\]
where $\tau \in [0,1]$ controls the update rate, providing stable training dynamics and encouraging convergence to consistent representations across different augmented views of the same image.

\subsubsection{BYOL training}
\label{sec:byol}

For \code{BYOL} training, we do not use the full $\sim$4 million BG images, as processing such a large dataset would be computationally prohibitive. Instead, we train on the $\sim$40,000  BGs in the G+24 sample, which is enriched in SGLs relative to their true occurrence, as it has been pre-selected by the G+24 SGL-finder and therefore contains a significantly higher fraction of lens candidates and lens-like systems than a random sample of the survey. This increases the prevalence of lens-like structures in the training data, allowing the model to more effectively learn features relevant to strong lensing, which would otherwise be extremely underrepresented.

In \citet{Mohale_2024}, building on \citet{Slijepcevic_2024}, the authors evaluate and adopt a set of data augmentation techniques (Table~\ref{tab:all_aug}). Their study focuses on a controlled morphology classification task involving well-defined and relatively abundant galaxy classes, allowing them to assess the impact of individual augmentations in isolation through a supervised downstream task. Their results show that optimal performance is achieved not by applying each augmentation separately, but by using a joint augmentation strategy that combines multiple augmentations during training.
The adopted scheme includes Gaussian blurring; horizontal and vertical flips (each with probability 0.5); random resized cropping (probability 0.7); and random rotations with angles $\theta \in [0,360]$, followed by cropping to 200 pixels and resizing to a standard output dimension. All augmentations are implemented using the \texttt{kornia} library \citep{Eriba_2019_kornia}.

We adopt the augmentation strategies of \citet{Mohale_2024} and split the dataset into training (70\%), validation (20\%), and downstream evaluation (10\%) subsets. The validation set is used to monitor training through the loss. Due to the strong class imbalance in our data, we are unable to reliably evaluate representation quality using a $k$-Nearest Neighbours classifier \citep[KNN,][]{Cover_knn}, as done in \citet{Mohale_2024}. As a result, we cannot systematically monitor or tune BYOL hyperparameters, including the choice and strength of augmentations, beyond tracking the validation loss. In contrast, \citet{Mohale_2024} are able to assess these effects in detail using KNN-based evaluation.

The PyTorch \citep{paszke_2019} implementation of \code{BYOL} (\texttt{BYOL-pytorch}) is publicly available on GitHub\footnote{\url{https://github.com/lucidrains/byol-pytorch}} \citep{Chen_2020_siamese}. The training hyperparameters are summarised in Table~\ref{tab:byol_hyp}. While \citet{Mohale_2024} report that 30 epochs are sufficient for convergence, we train for up to 50 epochs.
Early stopping with a patience of five epochs is applied, retaining the weights corresponding to the lowest validation loss. This strategy limits overfitting while allowing further refinement of the representations.
Finally, to assess training stability and reduce stochastic effects, \code{BYOL} is trained ten times with different random seeds. The model achieving the lowest validation loss is selected for downstream analysis.

\begin{table}
\centering
\begin{tabular}{lll}
\hline
Augmentation           & Parameters                                      & Prob. \\ \hline
Rotation with Crop        & Degrees $\in$ [0, 360], Crop Size: 200                        & 0.7                       \\
Vertical Flip             & -                                                      & 0.5                       \\ 
Horizontal Flip           & -                                                      & 0.5                       \\ 
Resized Crop              & Output Size: (244, 244), Scale $\in$ [0.7, 1]                 & 0.7                       \\ 
Gaussian Blur             & Kernel Size: (3, 3), Sigma $\in$ [1, 2]                       & 0.5                       \\ \hline
\end{tabular}
\caption{Data Augmentations Used for training \code{BYOL} with Corresponding Probabilities.}
\label{tab:all_aug}
\end{table}

\begin{table}
\centering
\begin{tabular}{ll}
\hline
Hyperparameter                       & Value    \\ \hline
Network architecture                 & ResNet18         \\ 
Epochs                               & 50                \\ 
Optimizer                            & Adam              \\ 
Learning rate                        & $1 \times 10^{-4}$ \\ 
Batch size                           & 64               \\ 
$\tau$ (EMA decay)                   & 0.99 (default)    \\
Hidden layer                         & avgpool           \\ \hline
\end{tabular}
\caption{Hyperparameters used in \code{BYOL} training}
\label{tab:byol_hyp}
\end{table}

\subsubsection{Feature extraction and dimensionality reduction}
\label{sec:feat_dim_red}

After \code{BYOL} training, the ResNet18 network is repurposed as a feature extractor. \citet{GUERIN_2021} showed that, for CNN-based clustering tasks, representations from the penultimate layer are the most informative. In ResNet18, this corresponds to the global average pooling layer (\texttt{avgpool}), which compresses the $64 \times 64 \times 3$ input into a 512-dimensional feature vector capturing high-level image representations. In the \texttt{BYOL-pytorch} implementation, the extraction layer can be adapted depending on the task, although \texttt{avgpool} is used by default.

Following the preprocessing steps of Sec. \ref{sec:preprocessing} the extracted features are then further reduced in dimensionality using PCA. PCA is selected over non-linear dimensionality reduction techniques, such as t-SNE or Uniform Manifold Approximation and Projection \citep[UMAP;][]{McInnes_2018}, following the findings of \citet{Mohale_2024}. While non-linear methods can introduce artificial clustering due to their flexibility, PCA better preserves global structure and reduces the risk of spurious clusters in feature space. UMAP is therefore used only for visualisation.

PCA is applied to reduce the dimensionality of the feature space. We test variance retention thresholds of 97\%, 98\%, and 99\%, and find that retaining 99\% of the variance provides the best performance on downstream tasks (Sec.~\ref{sec:control_sample}). We therefore adopt a 99\% threshold throughout this work.

\begin{figure*}
    \centering
    \includegraphics[width=0.7\textwidth, keepaspectratio]{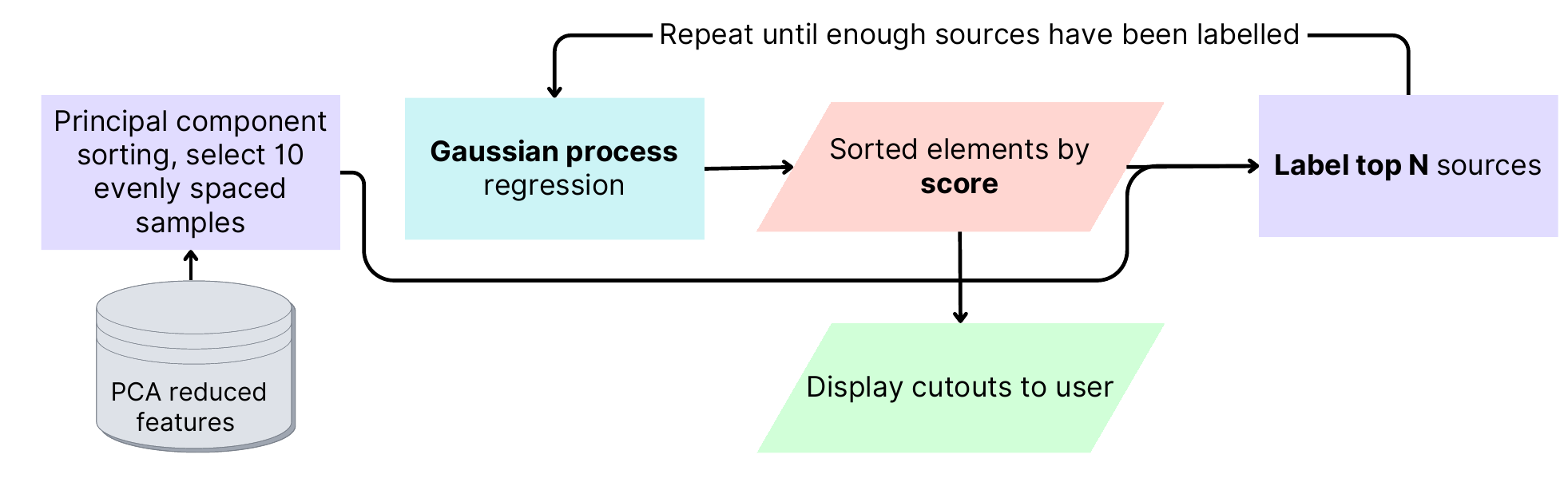}
    \caption{Workflow for the iterative labeling process with \code{Astronomaly}. The process repeats until the desired number of sources is labeled.}
    \label{fig:flow_labelling}
\end{figure*}

\subsection{SGL finding with \code{Astronomaly} }
\label{sec:SGLfinding_astro}
The \texttt{Astronomaly} framework comprises a Python back-end and a JavaScript front-end. The back-end performs data preprocessing and feature extraction, and implements anomaly detection and AL algorithms, while the front-end provides an interactive interface that allows users to explore the data and assign relevance scores to objects \citep{Lochner_2021}.

In this work, we use the AL loop and the front-end labelling interface provided by \texttt{Astronomaly}, while preprocessing and feature extraction are conducted independently. As this represents the first application of \texttt{Astronomaly} to SGL detection, we evaluate the feature-extraction stage separately to identify the most suitable approach. To this end, we develop a custom \code{BYOL} training pipeline following the methodology of \citet{Mohale_2024}.

The initial release of \texttt{Astronomaly} \citep{Lochner_2021} included two anomaly detection algorithms, Isolation Forest (iForest; \citealt{Liu_2008}) and Local Outlier Factor (LOF; \citealt{LOF_2000}), which are used to provide an initial ranking of the dataset and a starting point for AL. Feature extraction in that version was based on shape parameters, i.e., ellipse-fitting morphology.

The latest \texttt{Astronomaly} release, \code{PROTEGE} \citep{lochner_2024}, integrates BYOL for fine-tuning a feature extractor and supports a broader range of initial sorting strategies beyond purely anomaly-based approaches (see Sec.~\ref{sec:sorting}). It also introduces two distinct scoring mechanisms that incorporate user-interest–driven feedback, described in Sec.~\ref{sec:AL}.

A schematic overview of the \code{PROTEGE} AL workflow is shown in Fig.~\ref{fig:flow_labelling}.

\subsubsection{Initial Sorting}
\label{sec:sorting}

AL requires an initial set of examples to be presented to the user. Following the approach of \citet{Mohale_2024, lochner_2024}, data points in the PCA-reduced feature space are ranked, and ten sources are selected evenly along the first principal component to provide broad coverage of the feature space. These sources are presented to the user (or “oracle”) via the \texttt{Astronomaly} front-end, where they are scored based on perceived interest. The resulting labels are then used to guide subsequent learning.

Alternative strategies for initial sampling could also be considered. The \texttt{Astronomaly} framework, includes implementations of both LOF and iForest, and \citet{Verlon_2024} use these to show that LOF does not scale well to large datasets, whereas iForest remains effective for samples containing up to several million galaxies. However, such anomaly detection methods are inherently geared towards identifying outliers in low-density regions of feature space. This is at odds with the properties of SGLs in our evaluation sample, which, based on visual inspection, tend to lie in comparatively dense feature space regions. For this reason, we instead adopt the \code{PROTEGE} approach, as it is better suited to identifying rare sources even when they are embedded within dense regions of the feature space.

\subsubsection{Active Learning Loop}
\label{sec:AL}

Through the \texttt{Astronomaly} front-end, the user can iteratively assign labels and retrain the active learning model until they are satisfied with the results. The notion of an “interesting” source is intentionally defined by the user, allowing the AL to adapt to evolving scientific priorities and to target rare or unconventional systems.

In this work, scores between 0 and 5 reflect the perceived relevance for strong lensing: a score of 5 typically corresponds to a clear or highly promising SGL candidate, while lower scores are assigned to systems increasingly inconsistent with lensing (e.g. absence of blue arcs). These user-provided scores enable the algorithm to progressively prioritise sources exhibiting lens-like features. The grading scheme is not intended as a strict classification, but rather as a practical mechanism to down-weight obvious non-lenses and steer the active learning process toward regions of feature space enriched in potential strong lenses.

\begin{figure}
    \centering
    \includegraphics[width=0.8\linewidth]{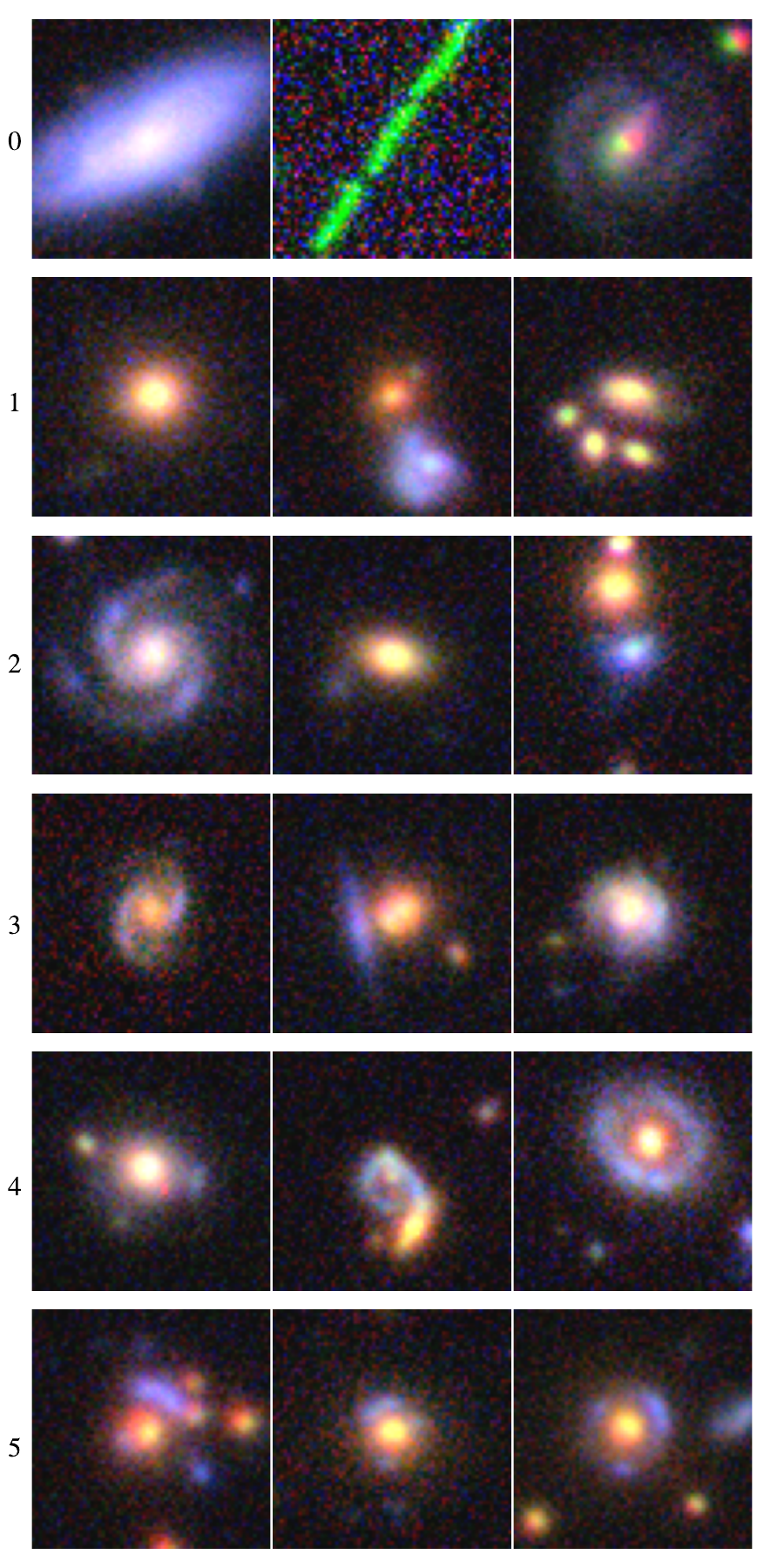}
    \caption{Example grades given during AL training.}
    \label{fig:astronomaly_grades}
\end{figure}

In the early stages, the user may intentionally assign higher scores to objects that only loosely resemble lensing systems in order to direct the model toward potentially relevant features. As the process progresses and the model becomes more sensitive to lens-specific characteristics, the grading naturally becomes more stringent, reflecting a refined understanding of the target population.
Examples of each grade are shown in Fig.~\ref{fig:astronomaly_grades}, and the criteria used to assign each grade:

\begin{itemize}
\item Grade 0: Images dominated by artifacts, nearby galaxies, or poorly subtracted sources.
\item Grade 1: Images showing no blue features or structures suggestive of lensing.
\item Grade 2: Systems containing a central galaxy that could in principle act as a lens (e.g. a red galaxy), or showing blue structures in the field, including features that may mimic lensing (e.g. spiral arms).
\item Grade 3: Objects exhibiting features suggestive of lensing (e.g. blue arcs or companion blue galaxies) that are unlikely to be true lenses, but whose morphology more closely resembles that of strong lenses than Grade 2 systems.
\item Grade 4: Systems that are not secure SGL candidates but display configurations consistent with lensing (e.g. an edge-on blue galaxy close to a central red galaxy or ring galaxies), with morphologies closely resembling those of strong lenses.
\item Grade 5: Secure SGL candidates.
\end{itemize}

For the evaluation tests (in Sec.~\ref{sec:control_sample}), conducted prior to application on the full DR4 dataset, we make use of the G+24 labels to explore different active learning configurations and parameter choices. In this controlled setting, labels are supplied directly to \code{Astronomaly}, bypassing the need for manual visual inspection and enabling reproducible assessment of the AL pipeline. These labels are used exclusively for testing; when applying the method to the full DR4 dataset, no such labels are available or used, and the process relies entirely on human-in-the-loop labelling.

\subsubsection{User interest prediction with Gaussian Processes}
\label{sec:GP}

Following the initial sorting, a GP is trained on the labeled feature vectors to model user scores, and is iteratively updated as new labels are provided.
The GP regressor is defined as
\begin{equation}
f(\mathbf{x}) \sim \mathcal{GP}\left(m(\mathbf{x}), k(\mathbf{x}, \mathbf{x'})\right),
\end{equation}
where $f(\mathbf{x})$ is a latent scoring function, $m(\mathbf{x})$ is the mean function (assumed to be zero), and $k(\mathbf{x}, \mathbf{x'})$ is the covariance function, implemented as the sum of a Matérn kernel and a white-noise kernel.

Given a set of labeled feature vectors, the GP is trained to infer the posterior distribution of $f(\mathbf{x})$. The trained score is defined as the posterior predictive mean, $\mu(\mathbf{x})$, which represents the model’s estimate of the user-assigned score for each object and is used to rank sources by their relevance or interest.

The acquisition score is instead computed using the Expected Improvement (EI) criterion \citep{mockus1991bayesian,jones1998efficient}, following the formulation adopted by \citet{Walmsley_2022_GP}:
\begin{equation}
\label{eq:EI}
EI(\mu, \sigma, x^*) = (\mu - x^* - \tau)\,\Phi(z) + \sigma\,\phi(z),
\end{equation}
with
\begin{equation}
z = \frac{\mu - x^* - \tau}{\sigma},
\end{equation}

where $x^*$ is the highest user score observed so far. The two terms in Eq.~\ref{eq:EI} correspond to exploitation and exploration respectively. The first term, $(\mu - x^* - \tau)\Phi(z)$, favours galaxies whose predicted mean score $\mu(x)$ exceeds the current best $x^*$, concentrating queries in regions of the BYOL embedding already known to be relevant. The second term, $\sigma\phi(z)$, favours galaxies in poorly sampled regions where the GP uncertainty $\sigma(x)$ is high, encouraging the AL to discover morphologically distinct populations elsewhere in the learned representation. Since the BYOL embedding is entirely unsupervised, such exploration is essential to avoid confining the search to a single neighbourhood of the feature space. The trade-off parameter $\tau$ controls the balance between the two: larger values suppress the exploitation term, promoting broader exploration, while smaller values concentrate labels near the current best-scoring region. As new labels are provided, the GP is retrained and both scores updated, allowing the AL to iteratively focus on the most interesting regions of the feature space.

As new labels are provided, the GP is retrained and both scores updated, allowing the AL to iteratively focus on the most interesting regions of the feature space. We evaluate sorting by trained and acquisition scores, and experiment with both the total number of labelled samples and the batch size at which labels are provided, assessing which configuration places true lenses higher in the ranked list.

\subsubsection{Performance Evaluation}
\label{sec:control_sample}

Within \texttt{Astronomaly}, several parameters can be tested, including the batch size of labelled objects before each AL training round ($n$), the score used to rank objects for labelling (trained score or acquisition score), and, when using the acquisition score, the exploration–exploitation trade-off parameter ($\tau$, see Eq. \ref{eq:EI}). We test $\tau$ values of 0.2, 0.5, 2, and 3  - larger values promote broader exploration of the feature space. 

To assess model performance, we calculate the number of SGL candidates recovered (recalled) within the ranked list ordered by the trained score, as a function of the number of inspected objects. Unlike supervised approaches, which compute recall over the full dataset, performance is evaluated progressively along the ranked list, focusing on the first $n$ objects prioritised by AL. This reflects the practical application of AL, where only a limited subset of top-ranked candidates is visually inspected. 
The AL scores range from 0 to 5. Objects with scores greater than or equal to three are treated as SGL candidates. As the AL stage relies on a single annotator (MG) and visual grading is inherently subjective, this threshold is adopted to avoid an overly restrictive selection at this stage. The selected candidates are subsequently inspected by three experts, and we consider as good candidates those assigned final grades A or B.

Once the optimal AL configuration is determined using the evaluation sample, it is applied to the full dataset. After the completion of the human-in-the-loop labelling used to iteratively retrain the AL model, the identified candidates are independently inspected by three experts. Objects assigned final grades A or B are then used as query systems for similarity search in the extracted feature space.

\subsection{Similarity search}
\label{sec:sim_search}

Our approach is conceptually similar to that presented in \citet{Stein_2022}. In that work, a SSL feature extractor is trained on 76 million images from the Dark Energy Spectroscopic Instrument Legacy Imaging Surveys \citep[DESI,][]{levi2019darkenergyspectroscopicinstrument}. Similarity between images is quantified using the cosine similarity of their representation vectors, and nearest neighbours (NN) are retrieved using Facebook AI Research’s \texttt{FAISS} library \citep{douze2025faiss}, which enables efficient large-scale similarity search.

Given the smaller dataset considered in this work, we compare FAISS with the \texttt{scikit-learn} implementation of $k$-NN, evaluating both Euclidean and cosine distance metrics. As no significant differences are observed between methods or metrics at this scale, we adopt KNN with a Euclidean distance.

While \citet{Stein_2022} demonstrate similarity search as a proof of concept using three Known lenses as queries, we apply this procedure systematically to all grade A and B candidates identified during the AL stage. In addition, a subset of grade C objects is randomly selected to assess the quality of their NN. For each query object, we retrieve the 512 NN in feature space, matching the choice adopted by \citet{Stein_2022} to enable a direct comparison. The similarity search is performed independently by the three experts, and all candidates identified through this procedure are subsequently re-evaluated to ensure consistency and robustness in the final classification.

The NN inspection interface developed for this work displays a mosaic of the closest neighbours and enables rapid grading through a click-based system (one click for grade A, two for grade B, and three for grade C). Previously identified or known lenses are highlighted, and the layout of the mosaic (number of rows and columns) can be adjusted dynamically. An image contrast control slider is also provided to facilitate visual assessment. By presenting candidates in a structured grid, this interface allows substantially faster screening than standard single-object visual inspection, enabling efficient skimming of morphologically similar systems.

\begin{figure}
    \centering
   \includegraphics[width=1\columnwidth]{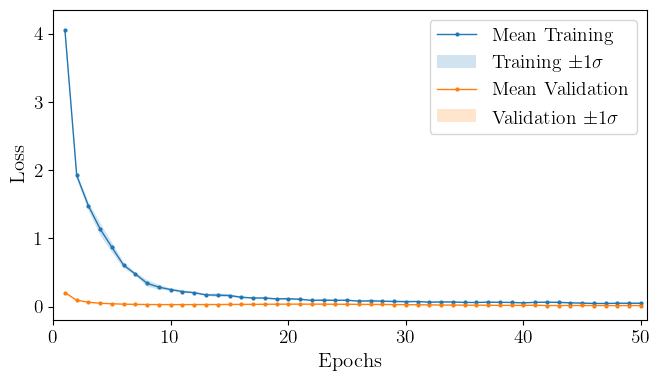}
    \caption{Average training and validation losses over 50 epochs, computed across ten independent runs. Shaded areas indicate the corresponding standard deviation.}
    \label{fig:loss_byol}
\end{figure}

\section{Results}
\label{sec:results}

\subsection{\texttt{BYOL} training}

The evolution of the BYOL training loss is shown in Fig.~\ref{fig:loss_byol}. The solid lines represent the mean over ten independent runs, and the shaded regions indicate the corresponding standard deviation. The training loss decreases rapidly during the first $\sim$10 epochs, followed by a more gradual decline, and approaches a plateau at later epochs. The validation loss converges more rapidly, stabilising after only a few epochs. The small and stable train-validation offset is expected in self-supervised training, where stochastic augmentations, batch normalization, and mini-batch optimisation introduce additional noise during training. The absence of divergence between the curves indicates stable optimisation and good generalisation of the learned representations. The mean duration of a complete 50-epoch run is approximately 14 h, corresponding to $\sim$17 minutes per epoch on a Tesla~P100-PCIE-12GB GPU.

The model achieving the lowest validation loss is used for downstream feature extraction. Applying PCA to the resulting embeddings shows that the G+24 BG sample requires 19 components to preserve 99\% of the variance, whereas the full KiDS DR4 BG sample requires 24 components to reach the same threshold. 
This difference is consistent with the distinct composition of the two samples. The G+24 sample has been pre-filtered by an SGL-finder and is therefore enriched in SGL candidates and SGL-like systems occupying a more restricted region of feature space, resulting in variance concentrated along fewer principal directions. In contrast, the full DR4 BG sample spans a broader range of galaxy morphologies and properties, producing a wider dispersion in feature space and hence a higher effective dimensionality.

\begin{figure*}
    \centering
    \begin{minipage}{\linewidth}
        \centering
        \includegraphics[width=0.62\linewidth]{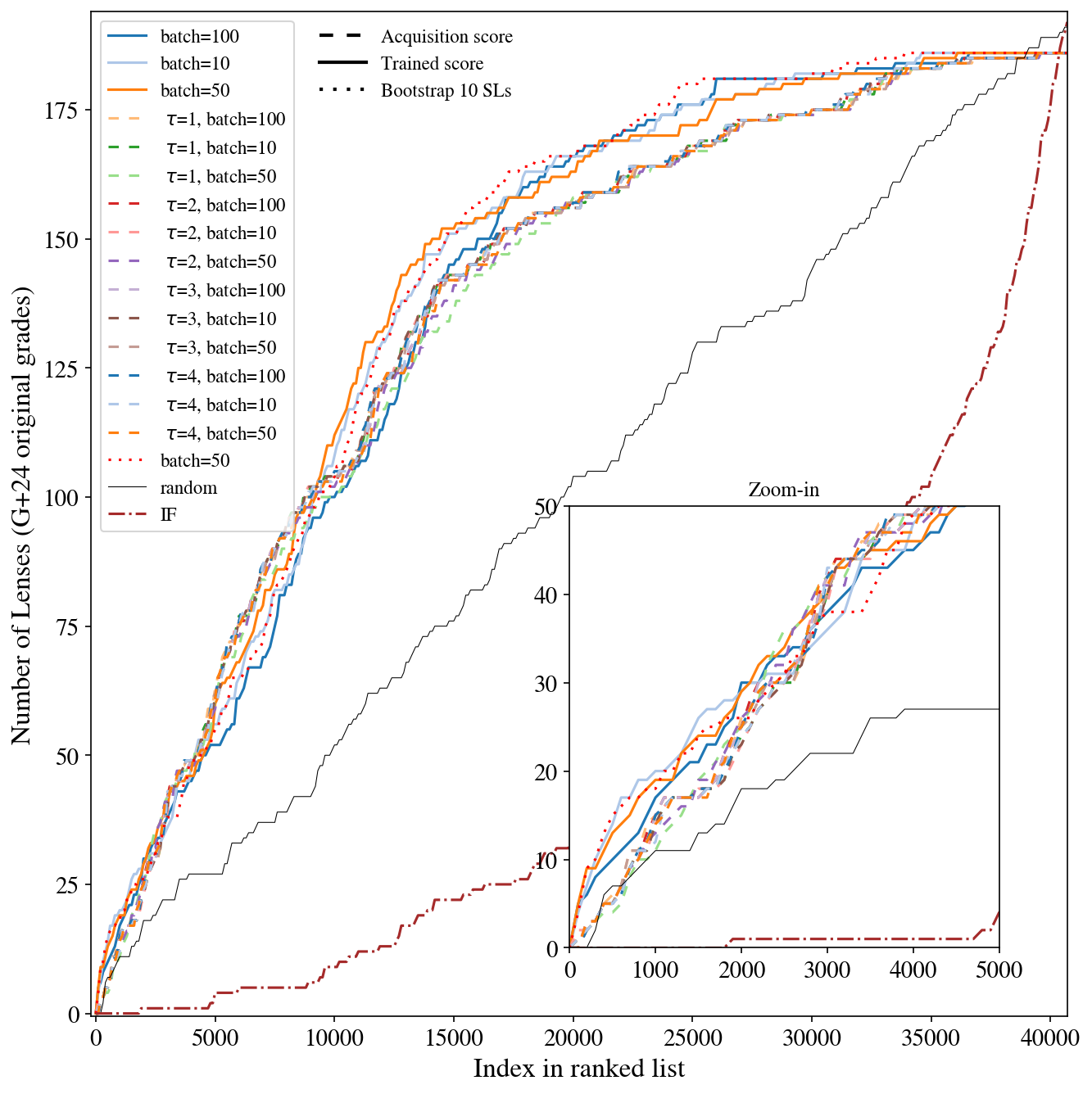}
    \end{minipage}

    \vspace{0.25cm}

    \begin{minipage}{\linewidth}
        \centering
        \includegraphics[width=0.62\linewidth]{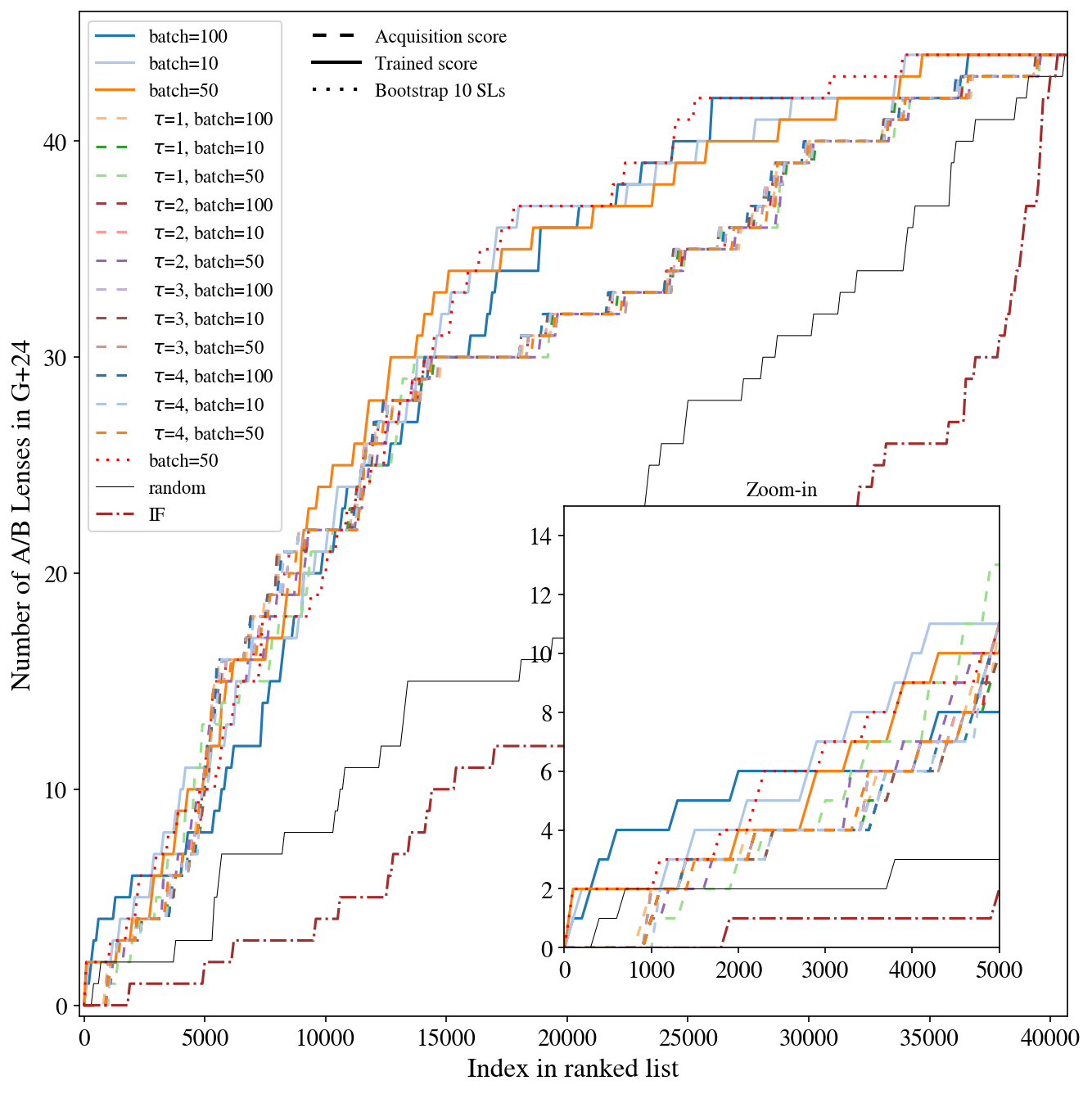}
    \end{minipage}

    \caption{Number of recovered SGLs as a function of rank position for AL runs trained with 2,000 labelled objects. The top panel uses the original G+24 classifications as ground truth, while the bottom panel adopts the re-inspected G+24 labels, considering only grade~A and B systems (Table~\ref{tab:origin_grades}) as SGL candidates. For comparison, random ordering and iForest (IF) ranking are also shown. All runs are initialised with 10 objects drawn from equally spaced regions of feature space (Sec.~\ref{sec:sorting}), except for one representative “bootstrap 10 SL” run initialised with 10 random known lenses (dotted curve). Different curves correspond to combinations of batch size, exploration parameter $\tau$, and scoring strategy; acquisition scores are shown with dashed lines and trained scores with solid lines.}
    \label{fig:recall_side_by_side}
\end{figure*}

\subsection{Evaluation set}

The \texttt{Astronomaly} AL regressor suggests to the user which objects to label next by assigning a score to each element in the dataset. Sorting the dataset by this score produces a ranked list that prioritises the most informative candidates for inspection.

Figure~\ref{fig:recall_side_by_side} shows the number of recovered SGL candidates as a function of rank position for AL runs trained with 2,000 labels, acquired iteratively in batches. Two ground-truth definitions are adopted and used consistently both to train the AL models and to evaluate the recovery performance.
The top panel adopts the original TEGLIE classifications, prior to the updated expert re-inspection in Sec.~\ref{sec:prev_disc}.
The bottom panel instead considers as true SGLs only the re-evaluated grade A and B systems in the G+24 dataset (Table~\ref{tab:origin_grades}). The labels of the 2,000 training objects are drawn from the corresponding ground-truth definition in each case.

Each curve corresponds to a different combination of hyperparameters: batch size, exploration parameter $\tau$, and sorting strategy. Results obtained using the acquisition score are shown with dashed lines, while those based on the trained score are shown with solid lines.
Within each sorting strategy, different colours denote different batch sizes, kept fixed throughout training until 2,000 objects are labelled. Only the best-performing configurations are shown; larger batch sizes (e.g. 150 and 500) yielded systematically lower recall and are omitted. For comparison, we show the expected behaviour under random ordering (solid black line) and the ranking obtained using Isolation Forest \citep[IF;][]{Liu_2008}. We also include a run initialised with 10 randomly selected known lenses (Bootstrap 10 SLs), whereas all other runs are initialised with 10 objects drawn from equally spaced regions of feature space (see Sec.~\ref{sec:sorting}).

The primary result is that the trained and acquisition scores yield comparable overall recall across all tested configurations, with no single strategy consistently dominating. This suggests that, once a sufficient number of labels has been provided, the GP predictions are reliable enough that both sorting criteria converge to similar rankings.
Focusing on the first few thousand ranked objects, corresponding to the number realistically inspectable by an expert (see inset), the trained score with batch = 100 and the bootstrap run perform marginally better in the top panel up to approximately 2,000 ranked objects, after which the acquisition score configurations begin to lead until 10,000 objects in the ranked list. In the bottom panel, batch = 100 leads in the early ranked objects.
Across all configurations, varying $\tau$ has a negligible impact on recovery, suggesting that the balance between exploration and exploitation in Eq.~\ref{eq:EI} is not a critical hyperparameter in this setting.

Bootstrapping with 10 example SGL candidates does not substantially improve performance, likely due to the sparsity and weak clustering of SGLs in feature space; the bootstrapped run performs comparably to, or slightly worse than, the equivalent configuration without bootstrapping. Both random ordering and IF exhibit substantially lower recall. The particularly poor performance of IF likely reflects its tendency to prioritise anomalous or low-density regions, which in this context are often dominated by artefacts or noise rather than genuine lenses, such that random sampling can in fact outperform anomaly-based ranking.

In the G+24 sample there are 50 or 200 SGL candidates, depending on whether the ground truth is taken before or after candidate re-inspection, in $\sim 4\times10^{4}$ objects. This corresponds to an SGL fraction of order $10^{-3}$, compared to the $\sim 10^{-4}$ expected in an unfiltered survey sample. As a result, the feature space explored here is enriched in SGL candidates, and hyperparameters such as $\tau$ may behave differently in a full-survey setting.
These tests should be interpreted primarily as a demonstration of the feasibility of applying AL to SGL discovery. The recall levels reported here are not intended to be representative of the performance expected on the full KiDS dataset.

\begin{figure}
    \centering
    \includegraphics[width=\linewidth]{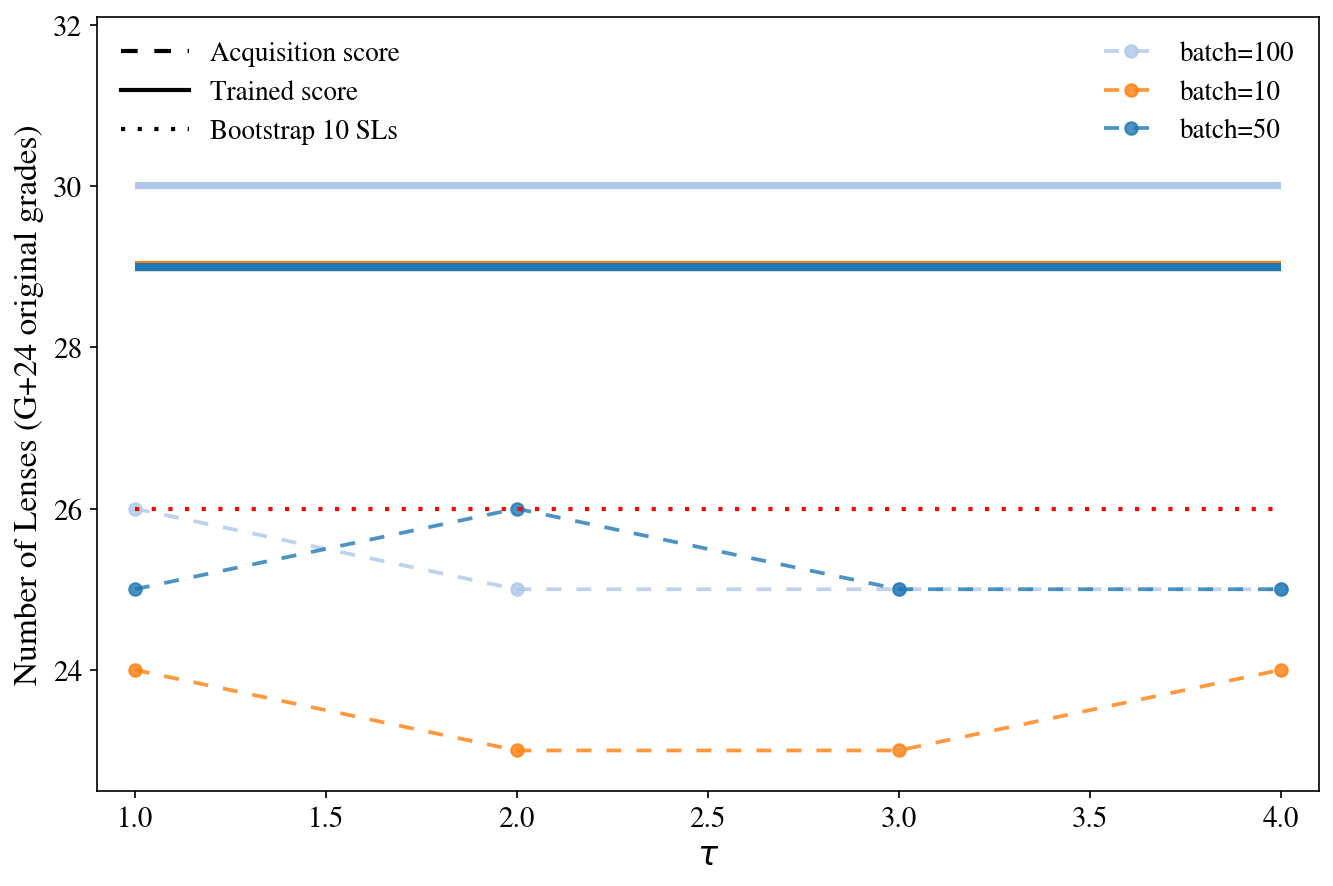}
    \caption{Performance at a fixed budget of 2,000 labelled objects, shown as a function of the exploration parameter $\tau$ using as ground truth the original G+24 labels.}
    \label{fig:tau_plot}
\end{figure}

To further clarify the behaviour in the early-inspection regime, we summarise the performance for the original G+24 labels at a fixed budget of 2,000 labelled objects in Fig.~\ref{fig:tau_plot}, where we show the number of recovered lenses as a function of the exploration parameter $\tau$. This provides an alternative representation of the highly crowded curves in Fig.~\ref{fig:recall_side_by_side}, isolating the dependence on $\tau$ at a fixed inspection depth. We find that the trained score consistently outperforms both the acquisition score and the bootstrap-10-SGL initialisation, reaching up to 30 recovered lenses (16.1\%) for batch size 100.

In contrast, for the stricter expert-defined A/B sample of this work (Table~\ref{tab:origin_grades}), all acquisition-based configurations yield identical performance at this stage, recovering 3 lenses (6.8\%) irrespective of $\tau$ and batch size (10, 50, and 100). The trained-score runs again achieve higher recovery, reaching 4 lenses (9.1\%) for batch sizes of 10 and 50, and up to 6 lenses (13.6\%) for batch size 100. The bootstrap initialisation recovers 4 lenses, performing comparably to the trained-score configuration with batch size 50.

Given the differing behaviours observed across the two evaluation sets, no single AL configuration can be considered universally optimal. As the AL stage adopts a less restrictive labelling threshold than that used in Table~\ref{tab:origin_grades}, we initialise the search using the acquisition score with a batch size of 10 and $\tau = 3$, thereby favouring exploration and maintaining fine-grained control over the early evolution of the ranking. This is particularly relevant given that only a few thousand objects will be inspected. Once the discovery rate decreases and the ranking becomes progressively dominated by previously explored regions of feature space, the strategy can transition to the trained score with larger batch sizes to favour exploitation and accelerate convergence.

\begin{figure*}
    \centering
    \includegraphics[width=\linewidth ]{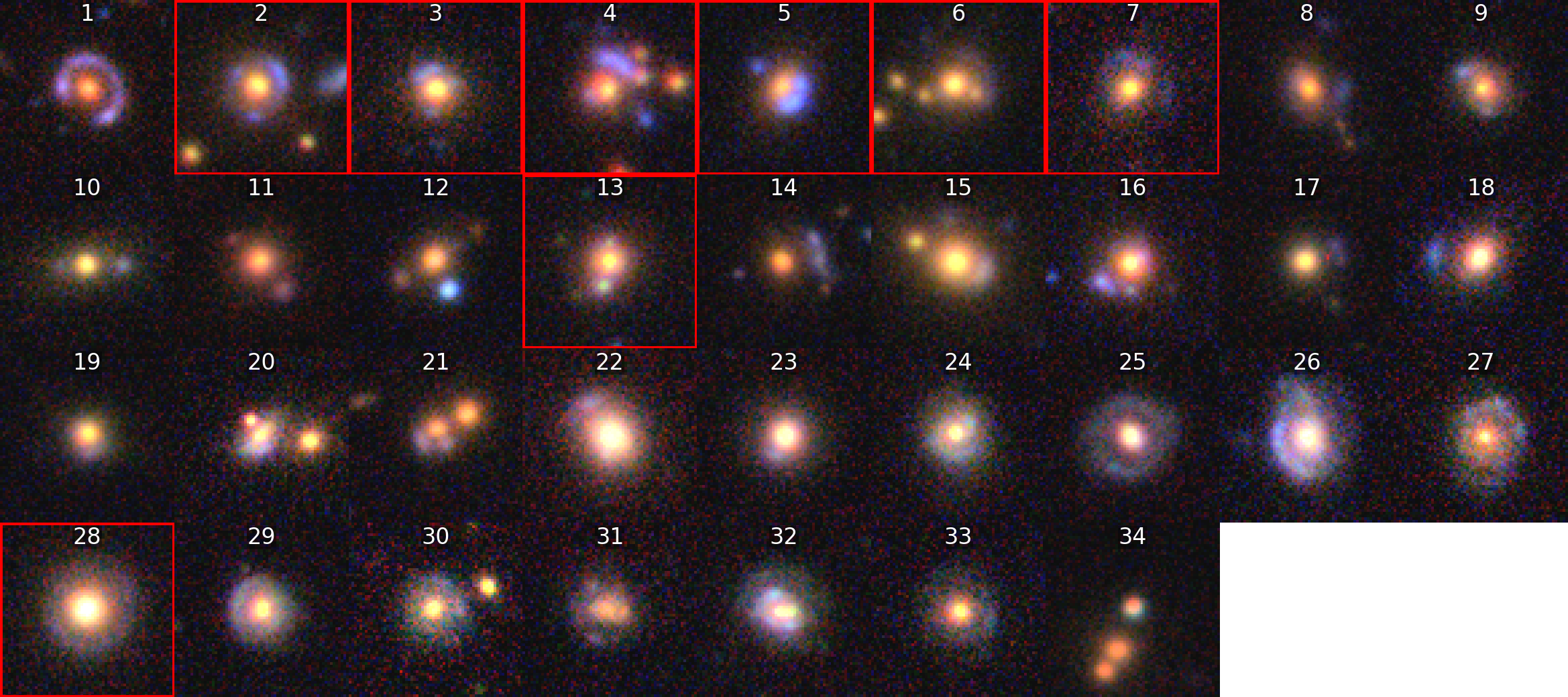}
    \caption{Mosaic of grade~A and B candidates identified during the AL training process. The five candidates (indices 1–5) correspond to grade~A systems, while the remaining panels show grade~B candidates. Red borders indicate objects that are part of the known lens sample. The indices displayed on each image correspond to their entries in Table~\ref{tab:AB_cand}.}
    \label{fig:AB_AL}
\end{figure*}

\subsection{Full KiDS DR4 sample}

Once the optimal AL configuration is established using the evaluation sample, we use the \texttt{Astronomaly} interface to visually label 3,000 objects from the 3,695,703 BGs present in KiDS DR4.

During the initial stage, the objects to label are selected using the acquisition score with $\tau$=3, which promotes exploration of different regions of the feature space. Every 10 labels we retrain the AL regressor and new images are suggested to the user. The early iterations are highly efficient, with an Einstein ring identified within the first 50 labelled objects. After $\sim$2000 labels, the discovery rate decreases noticeably, indicating that the most informative regions of the embedding have largely been sampled.
For the remaining 1,000 labels we switch to the trained score, which focuses on refining regions already identified as promising. This choice is motivated by two considerations: (i) from a methodological perspective, the subsequent similarity-search stage explicitly performs a local expansion around high-confidence systems in feature space. At this stage, further broad exploration through AL becomes less beneficial, as the most informative regions of the embedding have already been sampled. A similarity search therefore provides a more direct and computationally efficient way to probe nearby regions of feature space; (ii) from a practical perspective, the interactive labelling is performed remotely on a computing cluster. In this setup, each image must be loaded and rendered before grading, introducing latency between successive classifications.  As a result, the time required for large-scale manual inspection is dominated by image loading rather than by model retraining. In contrast, the similarity-search stage allows multiple neighbouring systems to be inspected within a single view, reducing the number of loading cycles and enabling more efficient expert screening.

The distribution of AL scores assigned during \texttt{Astronomaly} training (ranging from 0 to 5) is as follows: 0 (458 objects), 1 (840), 2 (937), 3 (469), 4 (213), and 5 (84). Adopting a threshold of score $\geq 3$ to define SGL candidates, a total of 766 objects are selected for further inspection by three experts. Of the inspected candidates, the final expert classifications are: 146 grade C, 29 grade B, and 5 grade A. This corresponds to confirmation rates (A+B+C) of approximately 15\% for AL score 3, 29\% for score 4, and 69\% for score 5, demonstrating a clear increase in purity with increasing AL score.

Figure~\ref{fig:AB_AL} presents a gallery of the 34 candidates identified during the AL search and assigned final grades A or B after expert inspection. The grade A systems are labelled from 1 to 5. Objects outlined in red correspond to systems already present in the `known' sample. The coordinates of these candidates are listed in Table~\ref{tab:AB_cand}, and the number shown in each stamp corresponds to the identifier used in the table.
All the candidates identified in this work are publicly available via Zenodo\footnote{\url{https://doi.org/10.5281/zenodo.21790924}}, which provides access to the grade A/B candidates as well as the complete list of grade C systems.

For visualisation purposes of the extracted feature space, we employ UMAP \citep{mcinnes_2018_umap}, a non-linear dimensionality reduction technique that preserves the large-scale structure of high-dimensional data but does not preserve exact metric distances. By projecting the features into two dimensions, UMAP provides a qualitative view of the how galaxies are organised in the embedding.

\begin{figure*}
    \centering
    \includegraphics[width=0.8\textwidth ]{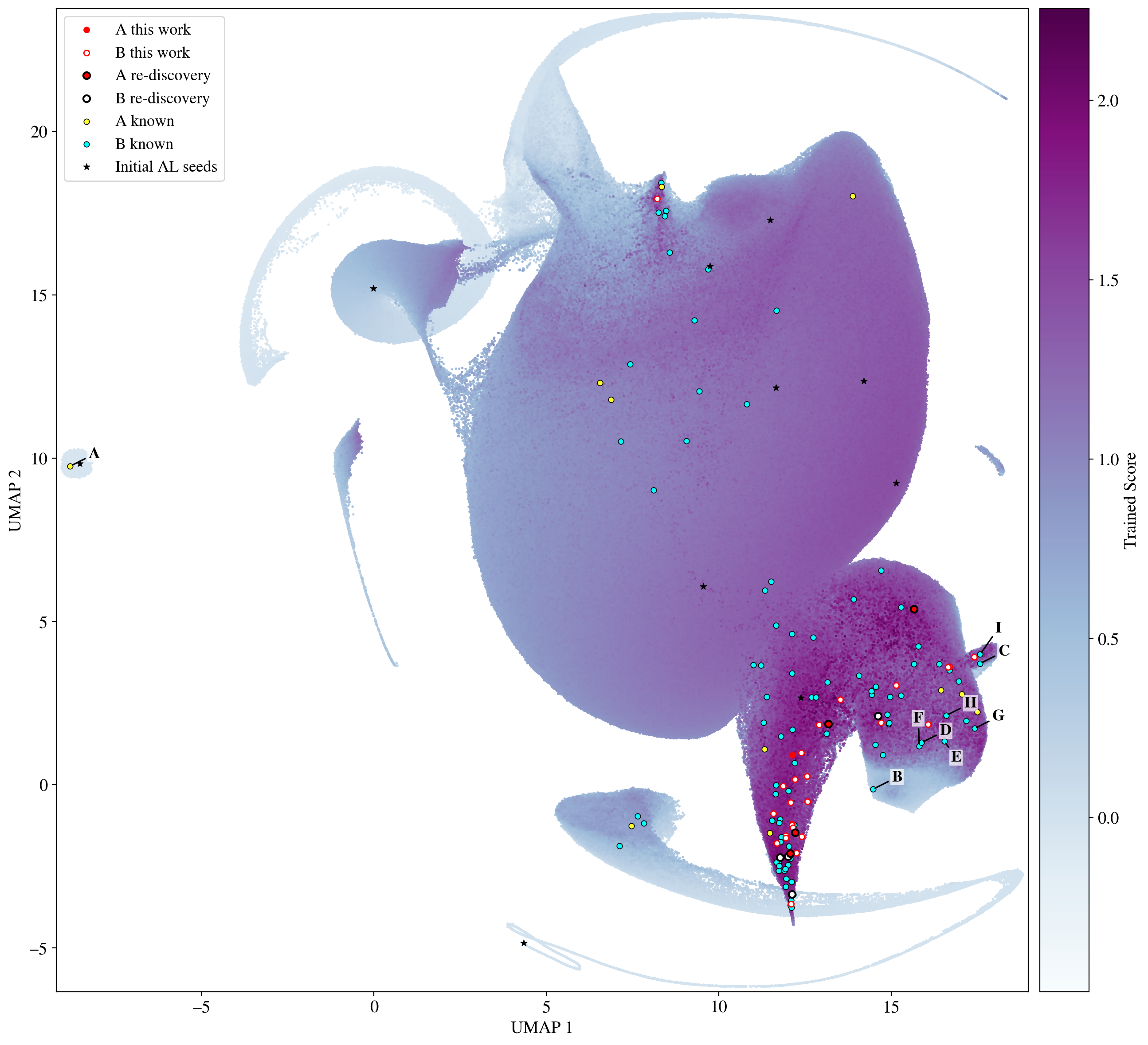}
    \caption{Two-dimensional UMAP projection of the 24-dimensional PCA-reduced feature representations for the full KiDS DR4 BG-filtered sample ($\sim$3.7 million galaxies). BG population is coloured by trained score (darker shades indicate higher values). Previously known SGL systems are shown in yellow (grade A) and cyan (grade B); newly identified candidates are marked in red (grade A) and gray (grade B) with red borders, and systems re-discovered during the AL stage are highlighted with yellow edges instead. Yellow stars denote the initial AL seed objects, i.e. the candidates used to initialise the first AL labelling round.
     The points associated with the most distant known lenses are labelled using the same letters as in Fig.~\ref{fig:far_known}.
    }
    \label{fig:umap}
\end{figure*}

Figure~\ref{fig:umap} shows the full sample of $\sim$3.7 million galaxies. Previously known SGL are overlaid, with grade A systems shown in yellow and grade B systems in cyan. New candidates identified in this work are marked in red (grade A) and white with red borders (grade B). Systems re-discovered during the AL training are plotted with the same colour scheme, but with larger markers and thicker black edges for emphasis. The BG population is coloured according to the trained score, with darker shades corresponding to higher values. 
A region with higher trained scores is visible in the embedding and the majority of newly identified and re-discovered candidates lie within or near this region. The 10 initial points selected as AL seed (black stars) are distributed across the embedding, reflecting the strategy of sampling from broadly separated regions of feature space at the start of training. Notably, the high-score concentration that emerges after training is not confined to the immediate neighbourhood of these initial seeds.This suggests that the AL model does not simply amplify the local surroundings of its starting points, but instead identifies specific regions of feature space associated with lens-like structure in the learned representation. Any remaining failures are therefore more plausibly attributed to limitations in the feature extraction: if lensing systems are not well captured by the adopted representation, they will not form a coherent region that can be reliably identified, irrespective of the exploration strategy.

\begin{figure}
    \centering
    \includegraphics[width=1\linewidth]{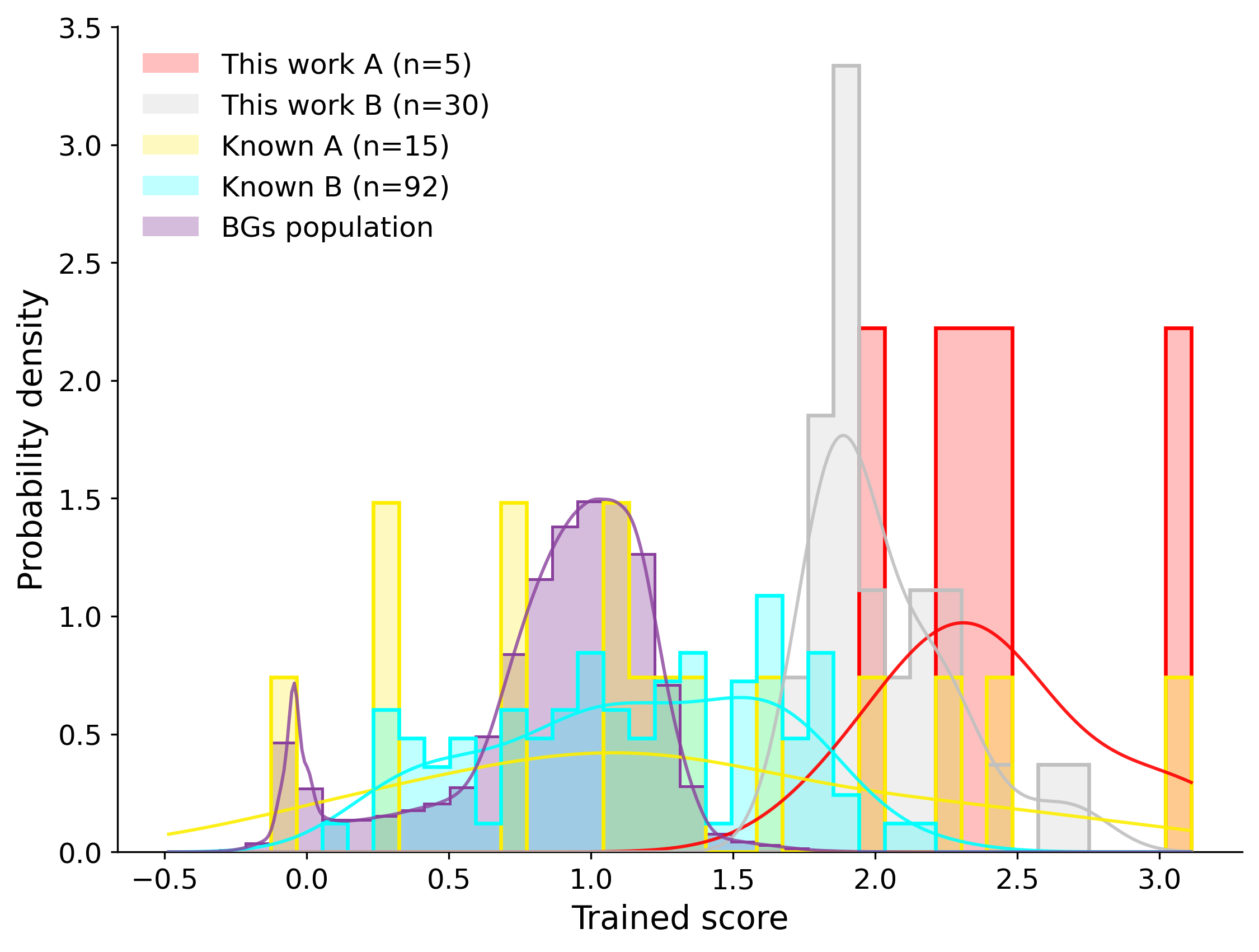}
    \caption{Probability density distribution of the trained score for the full BG population (purple), previously known SGL systems (grade A in yellow, grade B in cyan), and candidates identified in this work during the AL training (grade A in red, grade B in light red).}
    \label{fig:prob_desnsity}
\end{figure}

Fig.~\ref{fig:prob_desnsity} shows the distribution of the trained score after 3,000 labelled examples.
As expected, the newly identified SGL candidates exhibit higher trained scores than the BG population (median score $\sim 2.36$ for grade A systems, compared to $\sim 0.94$ for the full BGs sample). 
 In contrast, previously known lenses exhibit a broader range of scores, with substantial overlap with the BGs population. This overlap indicates that not all confirmed SGL systems are strongly prioritised by the final AL model. The fact that a fraction of confirmed lenses fall outside the highest-score percentiles likely reflects limitations in the data representation rather than the learning algorithm itself. In particular, it is driven by the preprocessing and feature extraction steps: if lenses occupy a coherent region in feature space, methods such as \code{PROTEGE} can effectively identify them, whereas systems that are not well captured by the chosen features are inherently difficult for any algorithm to recover.

To understand why 83 of the 107 known grade A/B SGL systems were not recovered during the AL stage, we examine their location in the learned feature space. For each known candidate, we compute its Euclidean distance to the set of known grade A/B systems recovered by AL. We then estimate the typical scale of this region by measuring the NN distances among the AL-recovered systems themselves (excluding self-matches), obtaining a median separation of $\sim$2.1 in feature space.

In contrast, known SGL candidates missed by AL have a median distance of $\sim$3.2 to the AL-recovered set, substantially larger than the internal scale of the recovered cluster. For comparison, the background population exhibits a median distance of $\sim$3.7.

These results indicate that the missed known SGL candidates do not lie within the compact region of feature space occupied by the AL-recovered lenses. Instead, they reside in geometrically distinct regions that are closer to the background population than to the core SGL manifold. This suggests that the incompleteness of the AL stage is primarily driven by a lack of sampling in those regions.

\begin{figure*}
    \centering
    \includegraphics[width=\linewidth]{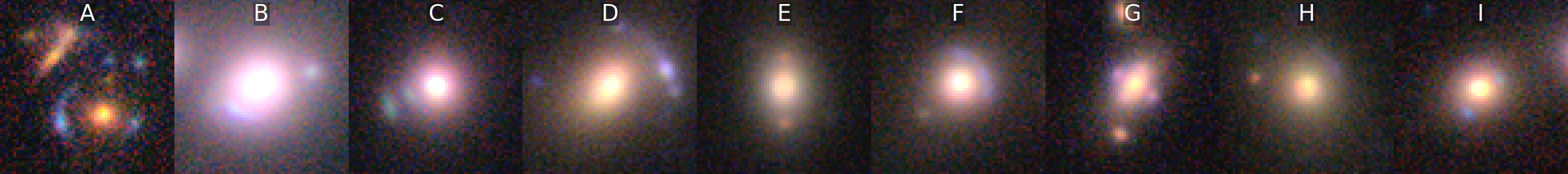}
    \caption{Examples of previously known strong gravitational lenses that were not recovered by our AL + NN search. The systems are shown ordered by decreasing distance from the grade A+B SGL manifold in feature space.}
    \label{fig:far_known}
\end{figure*}

To investigate whether certain regions of feature space remain unsampled by AL, we examine the known candidates that lie farthest from the AL-discovered SGL set. Figure~\ref{fig:far_known} shows the nine systems with the largest distances. The letters shown on each image correspond to the identifiers used to locate these systems in the UMAP projection shown in Fig.~\ref{fig:umap}. These objects exhibit a variety of morphologies, and no obvious visual trend explains why they lie geometrically far from the recovered candidates in feature space. 

In the UMAP projection, these systems lie at the periphery of, or outside (e.g. A and B), the high-score locus that contains most of the recovered candidates. Their apparent proximity to the main cluster in the two-dimensional embedding should therefore not be interpreted as evidence of similarity in the original feature space.

This non-detection can be explained in two ways: either the AL process did not sufficiently sample these regions of feature space due to the limited number of labelled images, or these regions were explored but do not contain SGL-like structures. To investigate this, we inspect the 512 nearest neighbours (NN) of each candidate shown in Fig.~\ref{fig:far_known}.

We first examine the preprocessed images used as input to the feature extractor. In one case (candidate A), the sigma-clipping step removes almost all signal from the image, leaving essentially no information for the feature extractor. This explains why the candidate is ranked poorly by the AL and represents the only instance where this occurs. As candidate A is not part of the G+24 dataset, this behaviour is not identified during testing. For the remaining candidates, the sigma-clipping preserves the lensing features; however, several systems exhibit relatively bright backgrounds, resulting in lower contrast between the lens and its surroundings, which may further affect the noise properties of the preprocessed images passed to the feature extractor.

For these systems, the corresponding neighbourhoods do not contain objects that would be classified as grade A or B lenses, with only a small number of grade C candidates identified across all nine systems. Instead, the NN appear to cluster primarily according to properties of the central galaxy rather than the lensing features themselves. In several cases, neighbouring objects are grouped by the morphology of the central source, while arc-like structures are not strongly reflected in the local feature-space similarity.
Understanding why some genuine SGL candidates occupy such regions of the embedding requires further investigation and may point to limitations in the current feature extraction stage. This could be explored using simulations in which the intrinsic properties of the systems (e.g. colours, magnitudes, and signal-to-noise ratios) are known, enabling a systematic assessment of how these properties propagate into the principal components and the resulting feature space.


\begin{figure*}
    \centering
    \includegraphics[width=\textwidth,height=0.97\textheight,keepaspectratio]{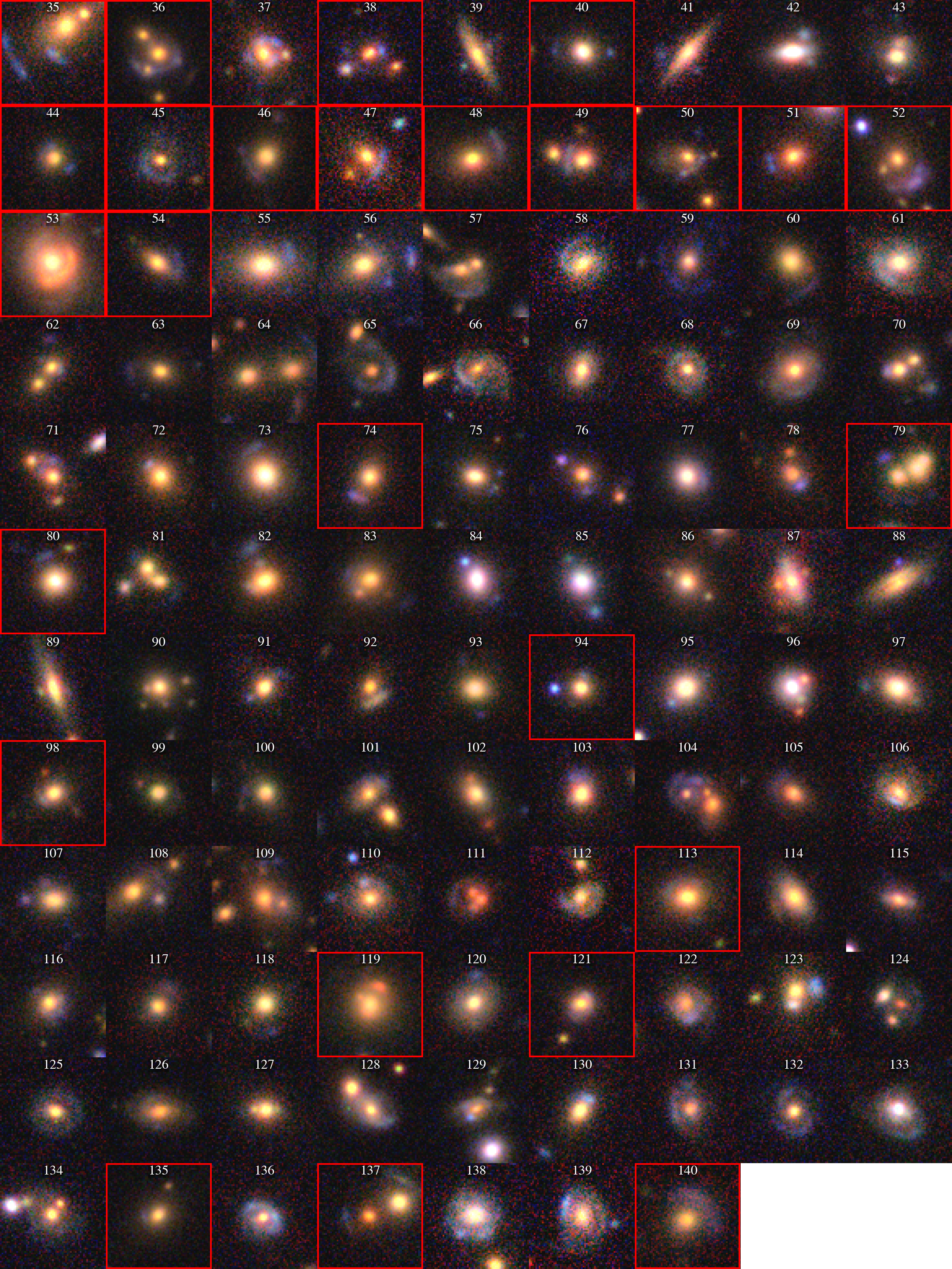}
    \caption{Mosaic of grade A and B candidates identified during the similarity search.
    Candidates 35, 36, and 37 correspond to grade~A systems, while the remaining panels show grade~B candidates.
    Cutouts highlighted with red borders correspond to the previously known SGL candidates.
    The indices displayed on each image correspond to their entries in Table~\ref{tab:AB_cand}.
    }
    \label{fig:AB_NN}
\end{figure*}

\begin{table*}
\centering
\caption{Grade A (marked with an asterisk $\ast$) and B SGL candidates identified in this work. Column “\#” matches the numbering in Figs.~\ref{fig:AB_AL} and \ref{fig:AB_NN}. KiDS ID (without “KiDSDR4” prefix), coordinates (deg), and literature references (where available) are listed. }
\label{tab:AB_cand}
  
\setlength{\tabcolsep}{2.8pt} 
\renewcommand{\arraystretch}{0.91} 

\begin{adjustbox}{max width=0.975\textwidth}
\begin{tabularx}{\textwidth}{
r l r r p{1.8cm}
r l r r p{1.8cm}
}
\toprule
\# & KiDS ID & RA & DEC & References & \# & KiDS ID & RA & DEC & References \\
\cmidrule(lr){1-5}\cmidrule(lr){6-10}
1*                    & J010127.840-334319.40   & 15.366003   & -33.722057  & [1], [2]     & 71                    & J021908.779-333119.81   & 34.78658    & -33.52217   &               \\
2*                    & J030639.688-273626.93   & 46.66537    & -27.607481  & [3]          & 72                    & J011153.430-314227.41   & 17.972626   & -31.707614  &               \\
3*                    & J013425.700-295652.42   & 23.607086   & -29.947897  & [4]          & 73                    & J142232.321+000134.55   & 215.634671  & 0.026264    & [7], [3], [5] \\
4*                    & J231645.663-292343.20   & 349.190266  & -29.395334  & [3]          & 74                    & J020505.970-331746.64   & 31.274878   & -33.29629   & [3]           \\
5*                    & J224546.032-295559.03   & 341.441801  & -29.933066  & [5]          & 75                    & J221647.408-292059.14   & 334.197537  & -29.349763  &               \\
6                     & J083726.163+015639.25   & 129.359015  & 1.944238    & [3], [6]     & 76                    & J000856.019-284610.08   & 2.233416    & -28.769469  &               \\
7                     & J145325.778-003331.75   & 223.357411  & -0.558822   & [7]          & 77                    & J030247.990-303627.51   & 45.699962   & -30.607643  &               \\
8                     & J234413.445-320057.62   & 356.056021  & -32.016007  &              & 78                    & J023300.432-330820.16   & 38.251804   & -33.138936  & [3], [2]      \\
9                     & J031432.791-254408.76   & 48.636633   & -25.735768  &              & 79                    & J230024.723-351233.67   & 345.103013  & -35.209355  & [3]           \\
10                    & J150421.290+001700.13   & 226.088711  & 0.283371    &              & 80                    & J091940.081-011630.76   & 139.917008  & -1.275213   & [5],[3]       \\
11                    & J032100.783-281810.47   & 50.253265   & -28.30291   &              & 81                    & J221016.394-344006.27   & 332.56831   & -34.668411  & [3]           \\
12                    & J000239.985-294635.77   & 0.666608    & -29.776603  &              & 82                    & J144503.938-032042.35   & 221.266411  & -3.345099   & [3]           \\
13                    & J001810.363-285609.54   & 4.54318     & -28.935984  & [4]          & 83                    & J005404.848-305809.32   & 13.520202   & -30.969256  & [10]          \\
14                    & J091523.562-005521.33   & 138.848177  & -0.922594   & [8], [5]     & 84                    & J122443.450-002811.36   & 186.181042  & -0.469824   &               \\
15                    & J000110.222-351533.95   & 0.292592    & -35.259432  &              & 85                    & J222209.601-335733.28   & 335.540008  & -33.959247  &               \\
16                    & J221322.958-285447.91   & 333.34566   & -28.91331   & [9]          & 86                    & J021932.170-332719.74   & 34.884042   & -33.455484  &               \\
17                    & J224144.452-282152.16   & 340.43522   & -28.364489  &              & 87                    & J115537.691+020815.48   & 178.907048  & 2.137635    &               \\
18                    & J111328.589+020948.55   & 168.369121  & 2.163488    &              & 88                    & J141325.378+010907.83   & 213.355743  & 1.152177    &               \\
19                    & J234450.820-314201.44   & 356.211752  & -31.700402  &              & 89                    & J031453.096-295233.96   & 48.721236   & -29.876102  &               \\
20                    & J141136.980+015357.13   & 212.904087  & 1.899205    &              & 90                    & J090150.268-012118.51   & 135.459453  & -1.355144   & [3],[5]       \\
21                    & J233323.587-322807.40   & 353.348281  & -32.468724  & [3]          & 91                    & J155300.545+001455.15   & 238.252274  & 0.248655    &               \\
22                    & J143001.826-015045.16   & 217.507611  & -1.84588    & [5], [3], [6]& 92                    & J001701.686-280827.57   & 4.257027    & -28.140993  &               \\
23                    & J142404.740-023019.89   & 216.019754  & -2.505526   &              & 93                    & J230150.841-352631.68   & 345.461839  & -35.442134  &               \\
24                    & J021110.350-302848.25   & 32.793128   & -30.48007   &              & 94                    & J230740.782-295058.93   & 346.919926  & -29.849703  & [5]           \\
25                    & J000837.779-330700.01   & 2.157415    & -33.116671  &              & 95                    & J151041.479-011903.79   & 227.672833  & -1.31772    &               \\
26                    & J113146.090+002212.70   & 172.942044  & 0.370195    &              & 96                    & J153136.971+002750.99   & 232.904048  & 0.464164    &               \\
27                    & J123752.030+000846.84   & 189.466793  & 0.146346    &              & 97                    & J114054.983-010757.27   & 175.229099  & -1.132575   & [5]           \\
28                    & J010642.065-342401.02   & 16.675274   & -34.400284  & [3]          & 98                    & J004204.610-323222.56   & 10.519212   & -32.539602  & [3]           \\
29                    & J001659.713-293119.84   & 4.248807    & -29.522179  &              & 99                    & J111433.617-020929.17   & 168.640074  & -2.158104   &               \\
30                    & J015059.897-290846.27   & 27.749573   & -29.146188  &              & 100                   & J015907.457-334415.29   & 29.781074   & -33.737582  &               \\
31                    & J111121.008-023548.31   & 167.837535  & -2.596755   &              & 101                   & J000017.928-353229.26   & 0.074701    & -35.541463  &               \\
32                    & J133708.148-000633.50   & 204.283952  & -0.109308   &              & 102                   & J000439.793-280507.82   & 1.165807    & -28.085506  &               \\
33                    & J000052.387-330330.92   & 0.218283    & -33.058589  &              & 103                   & J024126.819-284229.76   & 40.361747   & -28.708269  &               \\
34                    & J001625.370-313425.35   & 4.105712    & -31.573711  &              & 104                   & J032402.732-274857.72   & 51.011386   & -27.816035  &               \\
35*                   & J084520.257-005459.56   & 131.334408  & -0.916547   & [3],[5]      & 105                   & J024858.753-274354.53   & 42.244807   & -27.731814  &               \\
36*                   & J133844.788-010904.82   & 204.686617  & -1.151341   & [3]          & 106                   & J030016.828-321249.98   & 45.070119   & -32.213884  &               \\
37*                   & J235629.267-314515.77   & 359.121947  & -31.754382  &              & 107                   & J030511.509-311216.40   & 46.297955   & -31.204558  &               \\
38                    & J221400.330-292031.21   & 333.501378  & -29.342005  & [4]          & 108                   & J032523.797-324527.62   & 51.349158   & -32.757674  & [3]           \\
39                    & J021704.094-321155.44   & 34.267061   & -32.198735  &              & 109                   & J110051.121-005251.89   & 165.213006  & -0.881083   & [11]          \\
40                    & J115015.646-015752.32   & 177.565192  & -1.964534   & [5]          & 110                   & J033605.770-331220.58   & 54.024045   & -33.205718  & [3]           \\
41                    & J032621.242-345409.66   & 51.588511   & -34.902684  &              & 111                   & J223747.073-281546.48   & 339.446139  & -28.262912  &               \\
42                    & J025746.733-353418.53   & 44.444724   & -35.571814  &              & 112                   & J091321.371+025802.38   & 138.33905   & 2.967328    & [5]           \\
43                    & J031351.418-310215.66   & 48.464245   & -31.037684  &              & 113                   & J122450.300-004215.16   & 186.209584  & -0.704212   & [3]           \\
44                    & J123421.817-000925.99   & 188.590905  & -0.15722    & [7], [3]     & 114                   & J102729.477-013550.93   & 156.872823  & -1.597482   &               \\
45                    & J025334.181-284611.92   & 43.392423   & -28.769978  & [4]          & 115                   & J105741.640-004349.78   & 164.423502  & -0.730497   &               \\
46                    & J002105.099-283818.44   & 5.271248    & -28.638458  & [4]        & 116                   & J124215.591+020801.00   & 190.564966  & 2.133613    &               \\
47                    & J114444.815+001346.70   & 176.18673   & 0.22964     & [5],[3], [5] & 117                   & J135435.336-003719.64   & 208.647235  & -0.622124   & [3]           \\
48                    & J032219.773-342456.27   & 50.582391   & -34.415633  & [3]          & 118                   & J220606.000-351958.39   & 331.525002  & -35.332887  &               \\
49                    & J032230.223-344711.77   & 50.625931   & -34.786604  & [4]          & 119                   & J111555.997-025204.81   & 168.983322  & -2.868003   & [3]           \\
50                    & J131529.230+013223.65   & 198.871792  & 1.539903    & [3]          & 120                   & J140031.880+024610.26   & 210.132836  & 2.769518    &               \\
51                    & J141503.117-003105.93   & 213.762991  & -0.518314   & [5],[7]      & 121                   & J141649.819+013822.23   & 214.207581  & 1.63951     & [7], [5]      \\
52                    & J144950.700+005536.65   & 222.461254  & 0.92685     & [5], [7]     & 122                   & J145434.187-003428.19   & 223.642446  & -0.574499   &               \\
53                    & J083933.372-014044.81   & 129.889052  & -1.679115   & [7]          & 123                   & J235920.054-281535.45   & 359.83356   & -28.259849  &               \\
54                    & J232011.140-294158.79   & 350.046417  & -29.699664  & [5],[3]      & 124                   & J233430.820-313928.28   & 353.628417  & -31.657858  &               \\
55                    & J022816.291-292345.63   & 37.06788    & -29.39601   & [2]          & 125                   & J031950.805-295638.57   & 49.961688   & -29.944049  &               \\
56                    & J015116.013-290024.17   & 27.816722   & -29.006714  &              & 126                   & J223409.199-285423.36   & 338.53833   & -28.90649   &               \\
57                    & J033333.196-302441.95   & 53.388317   & -30.411653  &              & 127                   & J025825.003-264515.50   & 44.604181   & -26.754307  &               \\
58                    & J001941.102-332229.42   & 4.921261    & -33.374841  &              & 128                   & J231606.127-322814.63   & 349.025533  & -32.470731  &               \\
59                    & J154823.583+015832.68   & 237.098266  & 1.975745    &              & 129                   & J083734.985+024523.73   & 129.395775  & 2.756593    & [5]           \\
60                    & J141908.441+002049.65   & 214.785171  & 0.347127    & [7]          & 130                   & J023315.143-290354.57   & 38.3131     & -29.065159  &               \\
61                    & J013538.811-291531.46   & 23.911716   & -29.258741  &              & 131                   & J031554.060-264439.54   & 48.975251   & -26.744317  & [2]           \\
62                    & J141741.465+021705.13   & 214.422773  & 2.28476     &              & 132                   & J091903.838+003900.49   & 139.765994  & 0.650137    & [5]           \\
63                    & J002516.888-293016.99   & 6.320367    & -29.50472   & [3]          & 133                   & J121312.570+011853.41   & 183.302377  & 1.314837    &               \\
64                    & J013823.201-284408.09   & 24.596671   & -28.735581  & [2]          & 134                   & J122442.115-003959.04   & 186.175481  & -0.666401   &               \\
65                    & J235440.569-303219.81   & 358.669039  & -30.538838  &              & 135                   & J140733.603-014923.42   & 211.890016  & -1.823174   & [5]           \\
66                    & J151059.207+021938.68   & 227.746696  & 2.327413    &              & 136                   & J151037.311+024908.62   & 227.655466  & 2.819062    &               \\
67                    & J152408.309-014343.35   & 231.034621  & -1.728709   &              & 137                   & J152801.078+004030.62   & 232.004492  & 0.675174    & [3]           \\
68                    & J085247.537+001510.43   & 133.198073  & 0.252898    &              & 138                   & J220412.512-350654.12   & 331.052135  & -35.115035  &               \\
69                    & J150126.605-012501.29   & 225.360856  & -1.417026   & [11]         & 139                   & J222243.178-340745.35   & 335.679909  & -34.129265  &               \\
70                    & J234843.274-304922.09   & 357.180312  & -30.822803  &              & 140                   & J232039.461-281711.12   & 350.164421  & -28.286423  & [4]           \\
\bottomrule
    \end{tabularx}
\end{adjustbox}
\textit{Reference codes:}
[1]~\citet{bettinelli2016canarias};
[2]~\citet{Jacobs_2019}
[3]~\citet{links};
[4]~\citet{Li_2020};
[5]~\citet{grespan2024teglie}.
[6]~\citet{Sonnenfeld_2018_sugohi};
[7]~\citet{Jaelani_2020_sugohi}
[8]~\citet{Stein_2022}
[9]~\citet{Canameras_2020};
[10]~\citet{storfer_2024_new};
[11]~\citet{Huang_2021}
\end{table*}

\subsubsection{Similarity search}

The geometric analysis of the AL stage shows that the recovered SGL candidates occupy a coherent region of feature space. In this context, the NN stage acts as a local expansion of the SGL manifold identified by AL. While AL ranks objects according to a learned scoring function, NN retrieval directly probes the structure of the embedding. This motivates a complementary similarity-search stage aimed at exploring the local neighbourhood to identify additional candidates. For each grade A/B system and for 29 grade C objects (selected to match the number of grade B systems and ensure comparable statistics) identified during the AL stage, we retrieve their 512 NN in feature space using a Euclidean metric.

Table~\ref{tab:al_nn_completeness} summarises the recovery of previously known SGLs and the candidates identified in this work. “AL” and “NN” denote the number of candidates selected through the AL and NN stages, respectively, while “NN-only” lists candidates identified exclusively through the similarity search. “NN+AL” gives the total number of candidates when both stages are combined. “Recovered (AL)” and “Recovered (NN)” indicate the number of previously known lenses retrieved by each method. The final columns report the corresponding recovery fractions relative to the Known sample and the gain achieved when combining NN with AL, with “Gain from NN” representing the additional completeness obtained through similarity search beyond AL alone.

While AL alone preferentially recovers the highest-confidence systems (26.7\% of grade~A lenses), its recovery of intermediate-confidence candidates is limited, reaching only 4.3\% for grade~B and 1.5\% for grade~C systems. This behaviour can be understood in terms of global versus local search: the Gaussian Process underlying the AL framework learns a smooth, global function over the feature space, which favours regions associated with the most confident lens-like features but can miss more diverse or borderline cases. In contrast, the similarity search operates locally, identifying sources in the immediate neighbourhood of known candidates and thereby capturing a broader range of morphologies. As a result, the similarity search substantially improves completeness in these regimes, increasing recovery to 21.7\% for grade~B and 10.7\% for grade~C lenses. When combining AL and NN results, the overall recovery for grades A+B rises to 22.4\%, corresponding to a gain of 15 percentage points relative to AL alone.

The NN stage retrieves a large number of grade C (970) and grade B (127) candidates, far exceeding the numbers directly selected by AL. Among these, 886 grade C and 103 grade B systems are identified exclusively through similarity search and not during the human-in-the-loop AL training. This enrichment of intermediate-confidence systems suggests that the NN stage probes regions of feature space populated by morphologically consistent SGL-like structures.
Grade C candidates are individually less reliable, and most are unlikely to be genuine SGLs. Even so, they represent the largest part of both the known and newly discovered samples. The purity of the grade C sample cannot be established robustly, because doing so would require expensive follow-up with high-resolution imaging and spectroscopy. A rough expectation of 10-20\% is often assumed, although this depends on the image quality of the survey and other factors. For ground-based surveys such as KiDS, higher-resolution data from \textit{Euclid} or Roman would therefore be especially valuable. The key point is that the substantial number of grade C systems recovered by the similarity search indicates that the learned feature space consistently groups together objects with lens-like morphologies.

\begin{figure*}
    \centering
    \includegraphics[width=1\linewidth]{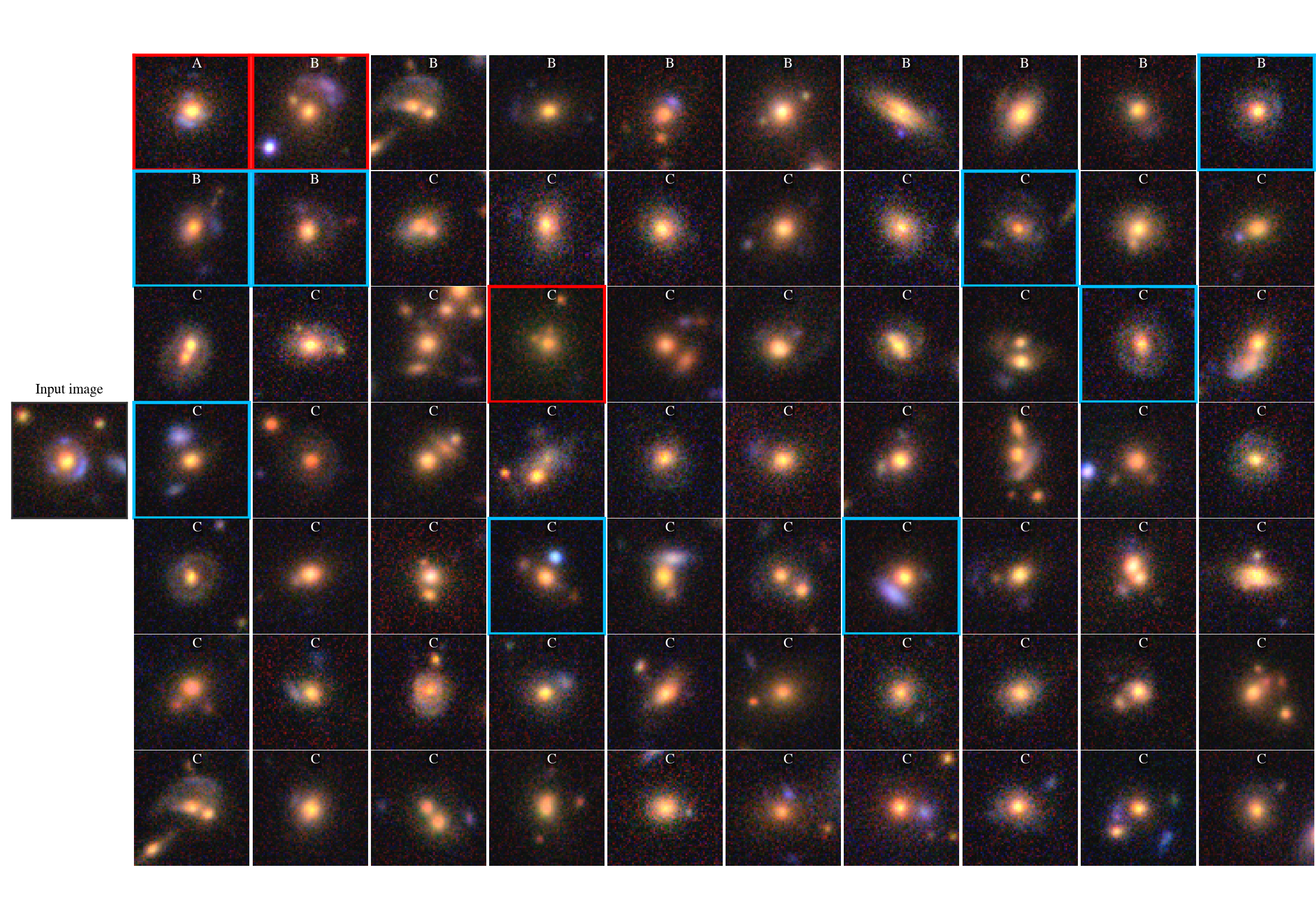}
    \caption{Example of a grade A candidate input image (index 2 in Fig.~\ref{tab:AB_cand}) and the candidates identified among the top 512 nearest neighbours on the right. Red borders indicate previously known systems that are part of the `known’ sample, while cyan borders mark candidates already identified during the AL stage. The final grade, assigned after visual inspection by three experts, is shown for each system. }
    \label{fig:query_nn_1} 
\end{figure*}

\begin{figure*}
    \centering
    \includegraphics[width=1\linewidth]{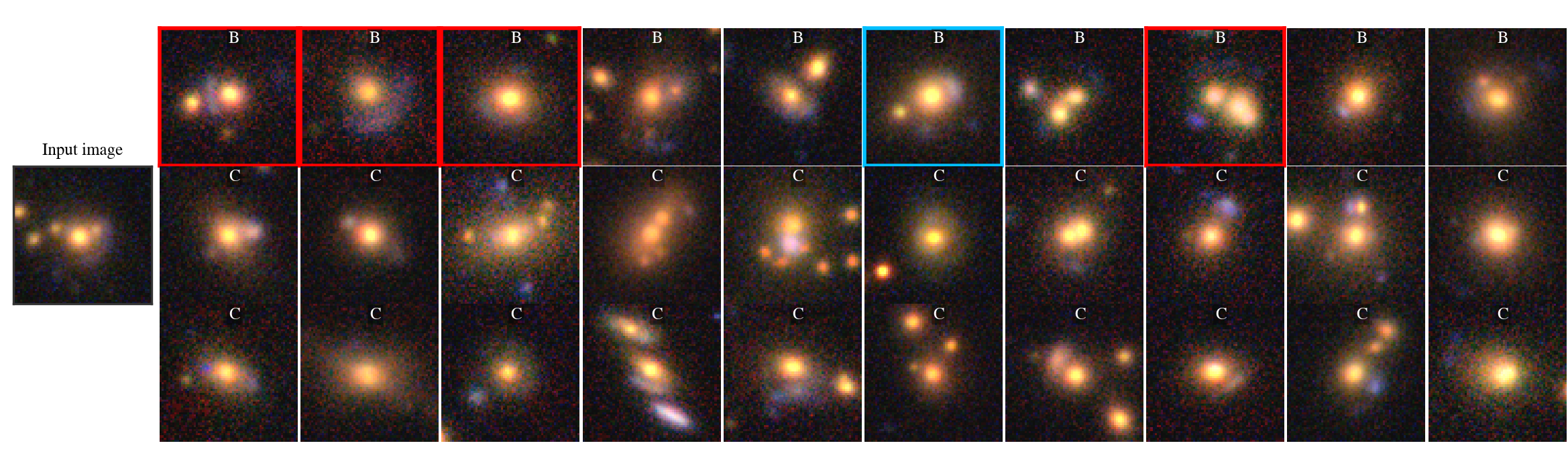}
    \caption{Example of a grade B candidate input image (index 6 in Fig. \ref{tab:AB_cand}) and the candidates identified among the top 512 nearest neighbours on the right. Red borders indicate previously known systems that are part of the `known’ sample, while cyan borders mark candidates already identified during the AL stage. The final grade, assigned after visual inspection by three experts, is shown for each system.}
    \label{fig:query_nn_2}
\end{figure*}

Two examples of SGL candidates and their selected NN are shown in Figs.~\ref{fig:query_nn_1} and~\ref{fig:query_nn_2}. In each case, the leftmost cutout shows an AL-selected candidate, while the surrounding stamps display its NN. Red borders indicate known SGL candidates, and blue borders mark systems previously identified during the AL stage.
Grade A systems exhibit, on average, a larger number of AB neighbours (7.3) than grade B (5.1) and grade C (2.4) systems, indicating that the highest-quality candidates preferentially reside in more densely populated regions of the embedding space.

\begin{table*}
\centering
\caption{
Number of previously discovered SGL candidates (`Known') from KiDS-specific searches and candidates identified in this work through the Active Learning (`AL') and Nearest Neighbour (`NN') stages, reported by grade. 
`AL' and `NN' give the total number of candidates selected by each method. 
`NN-Only' lists candidates identified exclusively through the NN stage. 
`NN+AL’ reports the total number of unique candidates obtained when both stages are combined.
`Recovered (AL)' and `Recovered (NN)' indicate the number of known systems rediscovered by each method. 
`Recovery AL' and `Recovery NN+AL' report the completeness relative to the Known sample, while `Gain from NN' gives the absolute increase in completeness when NN is added to AL.
}
\resizebox{\textwidth}{!}{%
\begin{tabular}{lcccccccccc}
\toprule
Grade
& Known
& AL
& NN
& NN-only
& NN+AL
& Recovered (AL)
& Recovered (NN)
& Recovery AL (\%)
& Recovery NN+AL (\%)
& Gain from NN (\%) \\
\midrule
A      & 15  & 5   & 5    & 3   & 8  & 4  & 0  & 26.7 & 26.7 & 0.0 \\
B      & 92  & 29  & 127  & 103 & 132  & 4 & 16 & 4.3  & 21.7 & 17.4 \\
C      & 206 & 146 & 970  & 886 & 1032 & 3 & 19 & 1.5  & 10.7 & 9.2 \\
A+B    & 107 & 34  & 132  & 106 & 140  & 8 & 16 & 7.5  & 22.4 & 15.0 \\
A+B+C  & 313 & 180 & 1102 & 992 & 1172 & 11 & 35 & 3.5  & 14.7 & 11.2 \\
\bottomrule
\end{tabular}
}
\label{tab:al_nn_completeness}
\end{table*}

\subsection{Cross-matching with existing catalogues}

Table \ref{tab:AB_cand} lists the 140 grade A and B SGL candidates identified in this work. The column “\#” corresponds to the numbering shown in the mosaics of Figs.~\ref{fig:AB_AL} and \ref{fig:AB_NN}. For each candidate, we report the KiDS ID (with the “KiDSDR4” prefix removed for brevity), the corresponding right ascension and declination in degrees, and references if system has been previously reported in the literature. 

To assess whether the identified candidates have been previously reported, we cross-match our sample against the SLED database\footnote{\url{https://sled.amnh.org/}}, the candidate list presented in G+24, and the compilation of SGL candidates used in G+24 the cross-matching. As expected, the largest overlap is found with SGL searches conducted within the KiDS survey itself, in particular the LiNKS \citep[][]{links} and  \citet{Li_2020} sample. Additional matches are found with the SuGOHI Hyper Suprime-Cam \citep[HSC,][]{Miyazaki_2012} samples \citep{Sonnenfeld_2018_sugohi, Sonnenfeld_2020_sugohi, Jaelani_2020_sugohi}, as well as catalogues derived from Dark Energy Survey \citep[DES,][]{DES_Abbott_2021} data \citep{Jacobs_2019, Huang_2021}, Pan-STARSS \citep{panstarrs_2012} survey \citep{Canameras_2020} and Dark Energy Spectroscopic Instrument \citep[DESI,][]{Decals_2019} \citep{Stein_2022, storfer_2024_new}.

Of the 140 grade A and B SGL candidates identified in this work, 81 have no counterpart in the consulted catalogues and are therefore considered new.\footnote{This literature cross-match is broader than the curated 107-object known-lens sample used to compute the AL/NN recovery statistics of Table~\ref{tab:al_nn_completeness}, which draws only on the LiNKS and TEGLIE samples; the two counts are therefore not directly comparable.}

Among the cross-matched systems, only one candidate, the Einstein ring reported with index 1, has been spectroscopically confirmed \citep{bettinelli2016canarias}.

\section{Discussion}
\label{sec:discussion}

In the context of astronomical imaging, self-supervised representation learning has been shown to extract meaningful morphological information directly from survey data without requiring extensive labelled training sets \citep[e.g.][]{bjerkeli2026astromorph, Walmsley_2022_GP}. Both contrastive and non-contrastive methods produce embeddings that preserve structural similarity and support downstream tasks such as morphology classification, clustering, and anomaly detection \citep[e.g.][]{Hayat_2021,Stein_2022,Mohale_2024,Slijepcevic_2024}. In particular, \citet{Mohale_2024} showed that BYOL-derived representations successfully group galaxies with similar morphology and enable unsupervised discovery workflows. Our results provide independent support for this picture: the enhanced neighbour density observed for SGL candidates indicates that lens-like morphologies occupy coherent regions of embedding space. This demonstrates that the learned representation captures meaningful structure relevant to SGL identification.

In total, we identify 140 grade A and B candidates using only 3,000 labelled galaxies. Although the recovery fraction of previously reported SGLs remains below 30\%, the remaining systems correspond to new candidates.

A more comprehensive analysis will be required to fully characterise the behaviour and limitations of the method. This includes assessing the contribution of individual features, the impact of colour scaling and preprocessing choices, and the expected gains from higher-resolution imaging. A rigorous evaluation needs to rely on realistic image simulations. However, publicly available simulations tailored to KiDS remain limited. The Bologna Lens Challenge simulations \citep{Metcalf_2019}, while informed by KiDS survey characteristics such as noise levels, are fully synthetic and do not use real KiDS images, instead generating mock systems with injected lensing features. As a result, they are not intended for training models to be directly applied to the survey and have been shown to differ significantly from real observations \citep{grespan2024teglie}. We therefore defer a comprehensive simulation-based assessment to future work on alternative datasets.

\subsection{Comparison with supervised approaches}

Supervised SGL finders typically achieve higher completeness, than unsupervised ones, when trained on large labelled datasets \citep[e.g.][]{Cheng_2020}. 
These models have been successfully applied to wide-area surveys \citep[e.g.][]{jaelani2023surveygravitationallylensedobjects, gonzalez_2025}, identifying hundreds to thousands of SGL candidates, and are now being deployed at scale in current surveys such as \textit{Euclid} \citep[e.g.][]{euclidcollaboration_2025_Walmsley}.

However, the performance of supervised models depends critically on the quality and representativeness of their training data, particularly when simulations are used. Mismatch between simulated and real survey data can introduce domain shift, which is a well known problem and has been shown to degrade performance through elevated false-positive rates or reduced recall when applied to real images \citep[e.g.,][]{grespan2024teglie, parul2025domain}.

In contrast, the AL + NN approach presented here requires no simulated SGLs and does not assume a predefined SGL class. Instead, it exploits the geometry of the learned representation space, guided by minimal expert feedback, to explore the data without strong reliance on training labels. The trade-off is clear: supervised methods tend to achieve higher completeness when domain alignment is strong and training labels are abundant, while AL + NN offers greater flexibility and robustness to domain shift, requiring fewer assumptions about the data distribution.

During the AL training, 34 grade A/B lenses are identified among 3,000 inspected objects, corresponding to a precision of $\sim$1.1\%. This is higher than the $\sim$0.6\% precision reported in G+24. When restricting the G+24 sample to the top 3,000 objects ranked by supervised model prediction probability, the precision decreases to $\sim$0.5\%, enabling a more direct comparison with the fixed inspection budget adopted in this work. For context, using the numbers reported in \citet{Li_2020} and \citet{Li_2021}, the supervised models achieve precisions of approximately 3.7\% and 7.9\%, respectively, when applied to similarly preprocessed KiDS BG samples.
It is important to note that \citet{Li_2021} trained their model on a substantially larger dataset of $\sim$80,000 images (including both mock lenses and non-lenses), which is not directly comparable to the more limited training set used in this work.

In \citet{gonzalez2025doesmachinelearningwork}, a comparative study of lens-finding models applied to DES data reports a maximum precision of 2.4\% for individual ML models. Notably, the best-performing model is trained using an AL strategy: the classifier is initially trained on simulated data, then applied to survey images and iteratively retrained using real examples to mitigate domain shift. In the same work, precisions exceeding 10\% are achieved only through ensemble combinations of multiple classifiers. These results highlight both the intrinsic difficulty of SGL discovery and the competitive performance of the label-efficient approach presented in this work.

\subsection{Active learning versus anomaly detection}

The large-scale \texttt{Astronomaly} application of \citet{Verlon_2024} showed that pure anomaly detection using iForest predominantly retrieves artefacts when applied to survey data, and that AL is required to prioritise scientifically meaningful outliers. Our results are consistent with this conclusion: blind anomaly detection alone does not efficiently recover SGLs, as illustrated by the iForest performance in Fig.~\ref{fig:recall_side_by_side}.

However, there is an important conceptual distinction between the two studies. \citet{Verlon_2024} aimed at broad anomaly discovery without a predefined target class. In contrast, our objective is the targeted discovery of SGLs. 

There are also methodological differences between the two studies. \citet{Verlon_2024} use the original \texttt{Astronomaly} framework \citep{lochner_2024} with features extracted using Zoobot \citep{Walmsley_2019}. In this work we instead employ \texttt{Astronomaly:Protege} \citep{lochner_2024}, which incorporates self-supervised representations learned with \texttt{BYOL}. Both studies analyse samples of comparable size (approximately four million objects), yet the earlier work identifies only eight SGL candidates among the first 2000 objects ranked by AL. This difference likely reflects the broader anomaly-driven objective of that study, which tends to prioritise less rare systems such as mergers in the ranking.

A further distinction lies in the initialisation strategy. Our search begins with ten equally spaced points in feature space (shown as stars in Fig.~\ref{fig:umap}), rather than seeding the process through trained scores. AL therefore does not simply re-rank statistical outliers, but instead guides exploration toward a specific morphological submanifold associated with lens-like configurations.

In \citet{oryan_2025_HST}, the active anomaly detection framework \texttt{AnomalyMatch} is applied to 99.6 million HST cutouts, identifying 140 candidate SGLs within the first 1,338 ranked objects. These candidates have not undergone detailed inspection by a dedicated expert team, and therefore the reported sample likely spans a range of confidence levels (e.g. grade A, B, and C candidates). \texttt{AnomalyMatch} represents an approach conceptually similar to the \texttt{Astronomaly} framework used in this work, as both aim to identify rare objects through self supervised representation learning and AL in large image datasets. However, while \texttt{Astronomaly} employs an active learning loop with human feedback to iteratively refine the ranking of interesting sources, \texttt{AnomalyMatch} relies on semi-supervised similarity matching to guide the search. A direct comparison or combination of the two pipelines could therefore provide an interesting avenue for future searches in large astronomical surveys.

In \citet{oryan_2025_HST}, approximately 60\% of the high-ranked anomalies had not previously been reported in the literature, highlighting the potential of active anomaly detection to uncover overlooked systems. Similarly, in our search we identify 140 grade A/B SGL candidates, of which 81 ($\sim$57\%) have no counterparts in existing strong lens catalogues. A dedicated comparison with supervised lens-finding methods would therefore be valuable to investigate which types of lens morphologies are preferentially identified by active learning and self-supervised approaches, and which may be missed by traditional supervised searches.

\section{Conclusions}
\label{sec:conclusions}

We present the first application of the \texttt{Astronomaly:PROTEGE}  AL framework to the discovery of SGLs. Feature representations are obtained using an ImageNet-pre-trained ResNet18 fine-tuned with the self-supervised \texttt{BYOL} pipeline. The search is conducted on 3.7 million bright galaxies in the KiDS DR4 survey selected to maximise the lensing cross-section.

We first investigate the impact of different AL training-loop configurations using a labelled sample of $\sim$40,000 BGs from \citet{grespan2024teglie}. To evaluate the effect of these hyperparameters, a consistent reference grading scheme is established. Three co-authors re-inspect all previously reported KiDS SGL candidates, and the same inspectors evaluate the candidates identified in this work to ensure uniform classification. 

We find that a more optimistic labelling schemes favour the acquisition score, whereas stricter grading strategies perform better when prioritising the trained score. We therefore adopt both scores: the acquisition score to promote exploration in the early stages, and the trained score focuses on higher-scoring regions of feature space. This setup is then applied to the $\sim$3.7 million KiDS DR4 images, with the AL stage trained using only 3,000 visually labelled objects ($\sim 0.1\%$ of the full dataset), resulting in the identification of 34 high-quality (grade A/B) SGL candidates.
Starting from these candidates, we perform a similarity search around the 5 grade A and 29 grade B systems, as well as around a subset of 29 grade C candidates. We identify 140 new grade A/B candidates and more than 1,000 grade C systems, and find that grade A candidates tend to have a larger number of similar neighbours in feature space. These results are enabled by the learned representations, which organise lens-like morphologies into coherent regions of feature space. By incorporating human labels, the AL stage concentrates inspection effort within these lens-enriched regions, while the similarity search probes the local metric structure of the embedding and enables the recovery of additional candidates. The similarity search significantly improves the recovery of known lenses, particularly for grade B and C systems, increasing the recovery fraction for A+B lenses from 7.5\% with AL alone to 22.4\% when NN candidates are included. Consequently, the final sample is dominated by newly identified candidates. 

At the same time, we find that previously known SGL candidates not recovered during the AL stage lie at significantly larger distances from the lens-rich region defined by the recovered systems. This may indicate that the current representation does not fully capture the diversity of SGL morphologies, or alternatively that the AL stage has not yet explored this part of feature space sufficiently. To investigate this further, we examine the nine known SGL candidates that lie furthest from the lens-rich region. Their local neighbourhoods contain no evident additional lens candidates, suggesting that the representation is dominated by features of the central galaxy rather than the lensed structures. 
Understanding which image components drive the representation requires datasets with known ground-truth lens and source properties. Simulated datasets therefore provide a way to determine how different morphological features contribute to the principal components of the learned feature space and to refine the resulting embedding, and will be explored in future work. 

Some limitations of the present approach should be acknowledged. 

We fine-tuned our feature extractor across ten independent BYOL runs to assess sensitivity to training stochasticity (Sec.~\ref{sec:byol}), but we did not carry out a full evaluation of how different runs affect the resulting feature spaces. The key question is whether they still keep similar objects close together. The selected run does this well, but a systematic comparison across runs is left for future work.

The AL training stage retains an element of subjectivity, as it relies on expert grading, which is inherently subjective and particularly challenging in the context of SGL identification. In particular, \citet{Rojas_2023} suggest that at least six inspectors are required for a stable classification. While this interaction is central to the methodology, it introduces variability typical of expert-driven searches. The impact of different labelling strategies has been introduced in \citet{lochner_2024}, but a systematic investigation of optimistic versus conservative labellers in the context of SGL detection would be valuable.

Future developments include improvements in representation learning, testing alternative feature extractors such as Zoobot \citep{Walmsley_2019}, incorporating multi-band or higher-resolution inputs. Surveys such as \textit{Euclid} and the Vera C. Rubin Observatory will produce datasets far larger than KiDS, making exhaustive human inspection infeasible. Current large-scale lens searches typically prioritise inspection of the highest-scoring candidates from supervised models. While this strategy efficiently reduces false positives, it may overlook rarer or atypical SGL configurations that are not well represented in the training data.

A full characterisation of the selection function of this method is also beyond the scope of the present work as it would require a different, more targeted design. We plan to carry out this analysis in future applications of \texttt{Astronomaly} to other surveys.

In this context, the framework presented here provides a flexible complement to fully supervised SGL finders, particularly in regimes where labelled data are limited or domain shift affects model performance. Because it operates directly on survey data and requires only minimal initial supervision, the method can adapt dynamically through human–machine interaction. It can generate representative real examples for supervised training, help mitigate domain shift, explore the neighbourhood of supervised detections, and identify configurations that specialised classifiers may miss. Used in tandem with supervised pipelines, such approaches can both validate high-confidence detections and probe regions of feature space that are less strongly prioritised by predictive models. Hybrid strategies of this kind may prove essential for maintaining discovery breadth in the era of billion-object surveys.

\section*{Acknowledgements}
MG acknowledges support from the Polish National Agency for Academic Exchange grant No. BPN/BEK/2024/1/00325/DEC/1. \\
ML, VE and KM acknowledge support from the South African Radio Astronomy Observatory and the National Research Foundation (NRF) towards this research. Opinions expressed and conclusions arrived at, are those of the authors and are not necessarily to be attributed to the NRF.
We acknowledge the use of the ilifu cloud computing facility – www.ilifu.ac.za, a partnership between the University of Cape Town, the University of the Western Cape, the University of Stellenbosch, Sol Plaatje University and the Cape Peninsula University of Technology. The Ilifu facility is supported by contributions from the Inter-University Institute for Data Intensive Astronomy (IDIA) -– a partnership between the University of Cape Town, the University of Pretoria, the University of the Western Cape, the Computational Biology division at UCT and the Data Intensive Research Initiative of South Africa (DIRISA).

ChatGPT (OpenAI) was used to assist with language refinement in this manuscript.

KiDS is based on observations made with ESO Telescopes at the La Silla Paranal Observatory under programme IDs 177.A-3016, 177.A-3017, 177.A-3018 and 179.A-2004, and on data products produced by the KiDS consortium. The KiDS production team acknowledges support from: Deutsche Forschungsgemeinschaft, ERC, NOVA and NWO-M grants; Target; the University of Padova, and the University Federico II (Naples).

\section*{Data Availability}

 The data underlying this article are based on the Kilo-Degree Survey Data Release 4 (KiDS DR4), which is publicly available on the KiDS website \url{https://kids.strw.leidenuniv.nl/DR4/}. The catalogues of candidates identified in this work are publicly available on Zenodo at \url{https://doi.org/10.5281/zenodo.21790924}.

\bibliographystyle{mnras}
\bibliography{mybibliography} 





\bsp	
\label{lastpage}
\end{document}